\documentclass[fleqn,usenatbib]{mnras}

\usepackage[T1]{fontenc}
\usepackage{ae,aecompl}

\usepackage{graphicx}	
\usepackage{amsmath}	
\usepackage{amssymb}	

\title[Tidal orbit decay among hot Jupiters]{Hierarchical Bayesian calibration of tidal orbit decay rates among hot Jupiters}

\author[A. Collier Cameron et al.]{
Andrew Collier Cameron,$^{1}$\thanks{E-mail: acc4@st-andrews.ac.uk (ACC)}
Moira Jardine$^{1}$
\\
$^{1}$Centre for Exoplanet Science, SUPA, School of Physics and Astronomy, University of St Andrews, North Haugh, St Andrews KY16 9SS, UK\\
}

\date{Accepted 2018 January 31. Received 2018 January 11; in original form 2017 September 11}

\pubyear{2017}

\begin{document}
\label{firstpage}
\pagerange{\pageref{firstpage}--\pageref{lastpage}}
\maketitle
%
\begin{abstract}
Transiting hot Jupiters occupy a wedge-shaped region in the mass ratio-orbital separation diagram. Its upper boundary is eroded by tidal spiral-in of massive, close-in planets and is sensitive to the stellar tidal dissipation parameter $Q_s'$. We develop a simple generative model of the orbital separation distribution of the known population of transiting hot Jupiters, subject to tidal orbital decay, XUV-driven evaporation and observational selection bias. From the joint likelihood of the observed orbital separations of hot Jupiters discovered in ground-based wide-field transit surveys, measured with respect to the hyperparameters of the underlying population model, we recover narrow posterior probability distributions for $Q_s'$ in two different tidal forcing frequency regimes. We validate the method using mock samples of transiting planets with known tidal parameters. We find that $Q_s'$ and its temperature dependence are retrieved reliably over five orders of magnitude in $Q_s'$. A large sample of hot Jupiters from small-aperture ground-based surveys yields $\log_{10} Q_s'=(8.26\pm 0.14)$ for 223 systems in the equilibrium-tide regime. We detect no significant dependence of $Q_s'$ on stellar effective temperature. A further 19 systems in the dynamical-tide regime yield $\log_{10} Q_s'=7.3\pm 0.4$, indicating stronger coupling. Detection probabilities for transiting planets at a given orbital separation scale inversely with the increase in their tidal migration rates since birth. The resulting bias towards younger systems explains why the surface gravities of hot Jupiters correlate with their host stars' chromospheric emission fluxes. We predict departures from a linear transit-timing ephemeris of less than 4 seconds for WASP-18 over a 20-year baseline. 

\end{abstract}




\section{Introduction}

The close-orbiting gas-giant planets known as ``hot Jupiters" are intrinsically very rare objects. They comprise a tiny minority of the population of transiting-planet candidates  detected in the course of the {\em Kepler} mission, yet they dominate the planet yield from wide-field, small-aperture ground-based surveys such as 
WASP \citep{2006PASP..118.1407P},
HAT/HAT-S \citep{2004PASP..116..266B,2013PASP..125..154B}
TrES \citep{2004ApJ...613L.153A},
XO \citep{2005PASP..117..783M}, 
QES \citep{2013AcA....63..465A},
KELT \citep{2007PASP..119..923P},
MASCARA \citep{2017A&A...601A..11T} and
WTS \citep{2012MNRAS.427.1877C}.
This is a natural consequence of observational selection: hot Jupiters are large planets with deep transits and short orbital periods, making them the easiest class of transiting planet to detect, and indeed the only class easily detectable by small ground-based instruments. The large areas of sky covered by these surveys have resulted in over 250 discoveries of such objects \citep{2011MNRAS.417.2166S}
\footnote{
http://www.astro.keele.ac.uk/jkt/tepcat/
}. 

Their Jupiter-like masses and short-period orbits make them ideal laboratories for studying  tidal interactions with their host stars. Periodic tidal deformation of an eccentric hot Jupiter dissipates energy in the planetary interior, first synchronising  the planet's rotation then circularising the orbit \citep{2004ApJ...610..477O,2010ApJ...725.1995M}. On a longer, and as-yet highly-uncertain timescale, the misaligned tidal bulges on the host star exert an unbalanced torque which transfers angular momentum from the planetary orbit to the stellar spin. Given long enough, this would lead to the planet's destruction: the most strongly-interacting hot Jupiters have insufficient orbital angular momentum to spin their stars up to Darwin-stable states in which both bodies rotate synchronously in a circular orbit \citep{2009ApJ...692L...9L,2015A&A...574A..39D}.

Detailed theoretical studies of tidal coupling to the stellar interior invoke two principal mechanisms: turbulent dissipation in the stellar convective zone \citep{2011ApJ...731...67P,2015A&A...580L...3M} and excitation of inertial wave modes in the radiative interior \citep{2007ApJ...661.1180O,2016ApJ...816...18E}. The strength of the coupling is parameterised in terms of a dimensionless tidal quality factor $Q_s'$, representing the inverse fractional energy dissipated per tidal forcing cycle.  Most hot Jupiters have very high forcing frequencies. Their late- F, G or K-type  host stars rotate significantly more slowly than their planets orbit, so the forcing frequency is close to the orbital frequency. In this case equilibrium tidal theory, dominated by turbulent dissipation, is applicable, leading to values of order $Q_s'=10^8$ \citep{2011ApJ...731...67P}. When the stellar spin frequency and the orbital frequency match within a factor 2 or so, the dynamical tide excites inertial waves in the radiative interior, decreasing $Q_s'$ and strengthening the rate of tidal orbit decay by up to three orders of magnitude \citep{2007ApJ...661.1180O}.

Direct attempts to measure secular period decreases in hot Jupiters through decade-long transit-timing campaigns have so far produced only upper limits \citep{2016MNRAS.455.1334H,2016AJ....151..137H,2017ApJ...836L..24W} or ambiguous detections that could be attributed to other phenomena such as apsidal precession in a mildly-eccentric orbit \citep{2016A&A...588L...6M,2016AJ....151...17J}. As the number of known hot Jupiters grows, however, statistical studies of the tidal erosion of the inner parts of the mass-separation (or mass-period) diagram \citep{2009ApJ...698.1357J,2010ApJ...723..285H,2011A&A...528A...2B,2012ApJ...757....6H} are gaining traction as a means of measuring $Q_s'$. Opinions differ, however on the values derived, with different studies delivering values ranging from $Q_s'\sim 10^6$ \citep{2016ApJ...820L...8M} to $Q_s'\sim 10^8$ \citep{2012ApJ...751...96P}. Some variation of $Q_s'$ from star to star is to be expected, since the rate of turbulent dissipation must depend in some way on the changing depth of convective zones along the main sequence. The expectation is that $Q_s'$ should increase towards higher effective temperatures if tidal dissipation rates increase with outer convective-zone depth \citep{2009MNRAS.395.2268B,2010MNRAS.404.1849B,2011ApJ...731...67P,2015A&A...580L...3M}.

Here we use Bayesian hierarchical inference to retrieve well-constrained numerical estimates of both $Q_s'$ and its dependence on stellar effective temperature, from the distribution of orbital distances of the known hot Jupiters from wide-field ground-based surveys.  In Section~\ref{sec:massep} we examine the physical and observational selection effects that determine the region within which transiting gas giants are found by ground-based, wide-field transit surveys. We employ the widely-used equilibrium-tide formulation of \cite {1998ApJ...499..853E} as implemented by \cite{2004ApJ...610..464D} as the generative model for the probability of finding a planet of given age at the observed separation from its star.

In Section~\ref{sec:mockcat} we simulate these processes, using a galactic model and equilibrium tidal theory to generate and evolve mock catalogues of planetary systems with known individual and global properties. We show that the input values of $Q_s'$ are recovered with better than factor-of-two uncertainty from mock samples whose size and observational and astrophysical selection effects are similar to those of the real hot Jupiters. We discuss the impact of our findings on several long-standing puzzles concerning hot Jupiters. These include the sensitivity of tidal coupling strength to stellar effective temperature, the apparent prevalence of rapid rotation and enhanced levels of  levels of chromospheric emission among stars hosting massive, close-orbiting  gas-giant planets, and the prospects for detecting tidal orbit decay through transit-timing variations.

\section{The mass-separation diagram}
\label{sec:massep}

The mass-separation diagram for transiting gas giants has a distinctive wedge-shaped appearance (Figure~\ref{fig:detection}), whose well-defined upper and lower boundaries have opposite slopes. The absence of both high and low-mass planets in short-period orbits cannot be attributed to observational selection effects. Planets of a given mass are easier to detect in short-period orbits, where they exhibit more frequent transits and their host stars' reflex orbits have greater velocity amplitudes. In this section we explore the observational and astrophysical selection effects that define the shape of the scatter plot of transiting gas giants in the mass-separation diagram.

\subsection{Tidal orbit decay}
\label{sec:orbdecay}

For a well-observed transiting system we have good estimates of the masses $M_s$, $M_p$ and radii $R_s$, $R_p$ of the host star and the planet respectively. 
The semi-major axis $a$ of a planet in a circular orbit about a slowly-rotating star decays at a rate defined by the equilibrium tide theory of \cite{1998ApJ...499..853E}, as implemented by \cite{2004ApJ...610..464D}:
\begin{equation}
\frac{1}{a}\frac{da}{dt}=-\frac{9I_s}{L_{\rm orb}}\left(\frac{n^2}{\alpha_s Q'_s}\right)\left(\frac{M_p}{M_s}\right)^2\left(\frac{R_s}{a}\right)^3.
\label{eq:dlnadt}
\end{equation}
We use the same definition as \cite{2004ApJ...610..464D} for the tidal quality factor $Q_s'$, with the tidal Love number incorporated. The orbital frequency $n$ is known. The stellar moment of inertia $I_s=\alpha_s M_s R_s^2$ and the effective squared radius of gyration $\alpha_s$ can be derived from stellar models, though since $\alpha_s$ cancels in eq.~\ref{eq:dlnadt} the model dependence disappears. The orbital angular momentum is $L_{\rm orb}=\lambda\sqrt{r}$ where $r\equiv a/R_s $ and 
\begin{equation}
\lambda\equiv M_p M_s\sqrt{\frac{G R_s}{M_p+M_s}}\simeq Mp\sqrt{G Ms Rs}.
\end{equation}
Kepler's third law gives $n^2=\eta^2 r^{-3}$ where
\begin{equation}
\eta^2\equiv\frac{G(M_p+M_s)}{R_s^3}.
\end{equation}
We thus obtain
\begin{equation}
\frac{1}{a}\frac{da}{dt}=\frac{9I_s}{\lambda }\left(\frac{\eta^2}{\alpha_s Q'_s}\right)q^2r^{-13/2}\equiv k r^{-13/2}
\label{eq:xrate}
\end{equation}
where $q\equiv M_p/M_s$ is assumed to be small, and 
\begin{equation}
k=\frac{9}{Q_s'}\frac{M_p}{M_s}\sqrt{\frac{G M_s}{R_s^3}}
\end{equation}
which has units s$^{-1}$.  Note in particular that $k$, and hence the rate of change of $\log r$, is inversely proportional to $Q_s'$. 

The time to spiral-in is estimated by rearranging eq.~\ref{eq:dlnadt} and integrating from $a=0$ to the planet's present position, to obtain 
\begin{equation}
t_{\rm remain}=\frac{2Q_s'}{117n}\frac{M_s}{M_p}\left(\frac{a}{R_s}\right)^5
\label{eq:spiralin}
\end{equation}
\citep{2011MNRAS.415..605B}.
Figure~\ref{fig:detection} shows the mass ratio - separation diagram for all planets listed in the Transiting Exoplanets Catalogue (TEPCat) of \cite{2011MNRAS.417.2166S} from the ground-based, small-aperture surveys of the WASP, HAT, TrES, XO, Qatar, KELT and MASCARA collaborations. The colour coding of the symbols indicates the spiral-in times for an assumed tidal quality factor $Q_s'=10^7$. 

It should be noted that eq.~\ref{eq:spiralin} for the spiral-in time $t_{\rm remain}$ tends to overestimate planetary life expectancy. The planet will either evaporate, undergo Roche overflow or impact the stellar photosphere on the way in. The time to migrate to the greatest of the three radii at which these things happen can be estimated using Eq.~\ref{eq:dr} below, and is incorporated in the more detailed analysis that follows.

\begin{figure}
\includegraphics[width=\columnwidth]{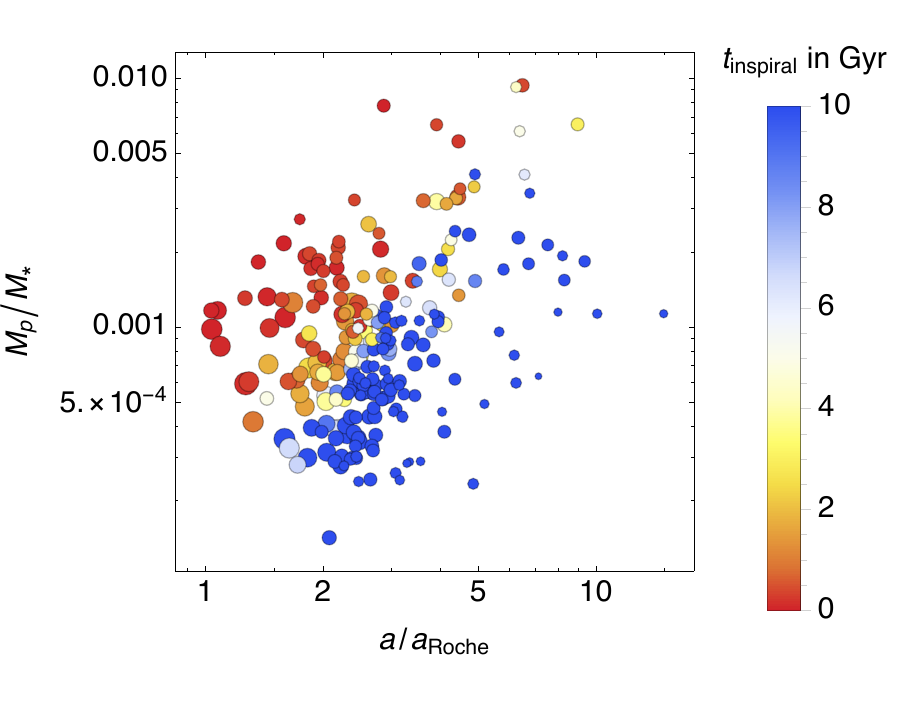}
    \caption{Planet/star mass ratio as a function of orbital separation in units of the planet's Roche radius $a_{\rm Roche}$. Symbol size is directly proportional to planet radius. Symbol colour denotes the planet's predicted life expectancy against tidal orbit decay, assuming $Q_s'=10^7$. Contours of constant life expectancy have the same slope as the upper-right boundary of the scatter plot. Planets along the upper-left boundary of the diagram have the shortest life expectancies.}
    \label{fig:mrocplot}
\end{figure}

\subsection{Evaporative erosion of low-mass, short-period planets}
\label{sec:evap}

\begin{figure}
\includegraphics[width=\columnwidth]{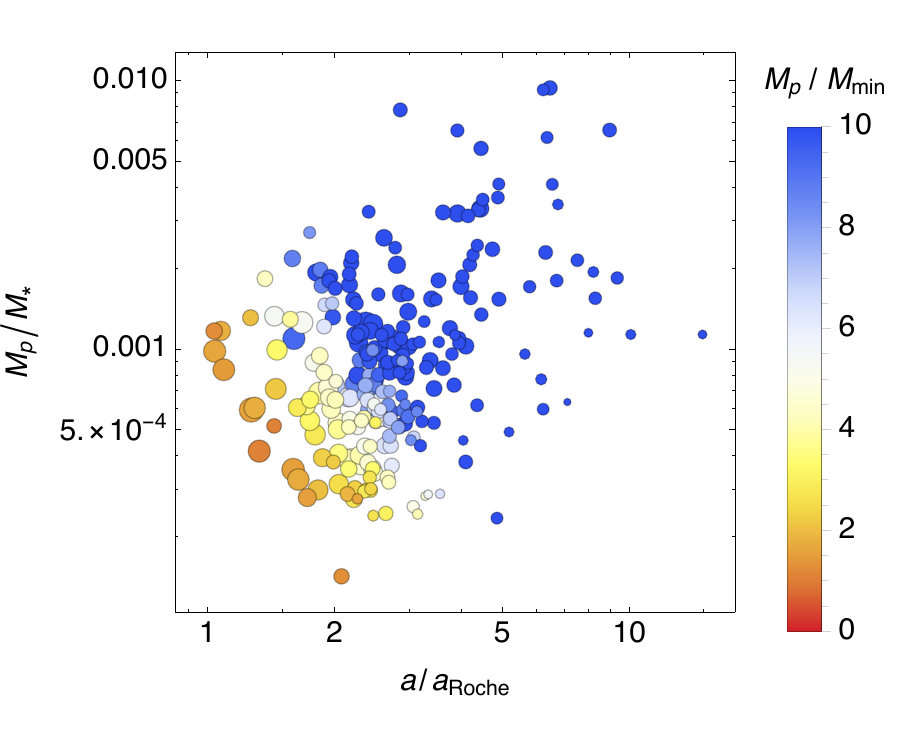}
    \caption{Planet/star mass ratio as a function of orbital separation in units of the planet's Roche radius $a_{\rm Roche}$. Symbol size is directly proportional to planet radius. Symbol colour denotes the ratio of the planet mass $M_p$ to the minimum mass $M_{\rm min}=(a_{\rm Roche}/a)^{2.5}$ at the inner boundary for a given separation.}
    \label{fig:KHplot}
\end{figure}

As the number of transiting planet discoveries has grown, the lack of low-mass, short-period hot Jupiters and Neptunes, first  noted by \cite{2005MNRAS.356..955M} has become more starkly apparent \citep{2012MNRAS.422.3151H,2016A&A...589A..75M}. \cite{2006ApJ...638L..45F} showed that the gas-giant planets known at the time all lay at separations greater than twice their respective Roche limits: 
\begin{equation}
a_{\rm Roche}=2.46 R_p \left(\frac{M_s}{M_p}\right)^{1/3},
\end{equation}
and pointed out that eccentric migration followed by tidal circularisation would produce a dearth of planets with $a<2a_{\rm Roche}$ and a pile-up near $a=3a_{\rm Roche}$.

When the presently-known sample is plotted as mass-ratio versus orbital separation in units of the Roche limit, it becomes apparent that, contrary to expectation from eccentric migration theory, many gas giants orbit well inside twice their Roche limits (Fig.~\ref{fig:KHplot}). Unless tidal orbit decay has carried them closer to the star within the system's lifetime, Type II migration appears to be a more likely means of transporting them to their present locations. 

The sharp lower-left boundary of the planet distribution in this diagram can instead be understood in terms of erosion by XUV irradiation. 
\cite{2004A&A...419L..13B} suggested that when a planet is so strongly irradiated by stellar XUV radiation that its mass-loss timescale $t_{\rm evap}\equiv M_s/\dot{M_s}$ becomes ten or so times shorter than the Kelvin-Helmholtz timescale of the planet, the planet would undergo catastrophic hydrodynamic expansion leading to the loss of all or most of its hydrogen-rich envelope, leaving only a stripped core.  Subsequent studies of XUV-driven mass loss by \cite{2007ApJ...658L..59H,2007Icar..187..358H} based on the work of \cite{1981Icar...48..150W} arrived at mass-loss rates two orders of magnitude less than those used by Baraffe et al. \cite{2006Icar..183..508Y}, in a corrigendum to an earlier paper \citep{2004Icar..170..167Y} had arrived at a rate intermediate between the Watson and Baraffe rates.

These energy-limited mass-loss schemes are valid only for low UV fluxes. In this regime, the UV photon energy serves to lift material out of the planet's potential well. At higher fluxes, however, the radiative loss function of the gas serves as a thermostat, keeping the gas temperature at about $10^4$ K, restricting the mass-loss rate in a hydrodynamic, transonic wind to $\dot{M} = 4 \pi \rho_s c_s r_s^2$ \citep{2009ApJ...693...23M}. Here $\rho_s$ and $c_s$ are the density and sound speed at the sonic point $r_s$.

\cite{2014ApJ...783...54K} included this radiation-recombination limitation in a population-synthesis study of the inner boundary of the hot Jupiter population. They concluded that runaway atmospheric escape and Roche-lobe overflow are expected to deplete the inner edge of the hot Jupiter population for core masses less than 10 Earth masses, providing a likely explanation for the dearth of sub-Jovian giant planets at small distances from their host stars.

In the present study, we need to be able to to describe the location of the inner boundary of the sub-Jovian planet population in at least an approximate way, as part of the generative model for the probability density function. It is difficult to assess the proximity of an individual giant planet to the evaporation boundary, without knowing its core mass. The sculpting effects noted by Kurokawa et al only apply to planets considerably less massive than Jupiter and involve Roche overflow. For simplicity, we choose to impose an empirical boundary at a mass-dependent multiple of the Roche limit. Fig.~\ref{fig:KHplot} shows that the inner boundary of the distribution is indeed close to the Roche limit for Jupiter-mass planets, and at twice the Roche limit for planets of $0.1 M_{\rm Jup}$. The minimum mass for a given value of $a/a_{\rm Roche}$ is therefore set at $M_{\rm min}=(a_{\rm Roche}/a)^\eta$ Jupiter masses, where $\eta\simeq 2.5$, for the purposes of the present study.

\subsection{Transit detection threshold}

Ground-based surveys such as WASP typically achieve an RMS scatter of order 0.01 magnitude per 30-s exposure at magnitude $V=12$ \citep{2004PASP..116..266B,2006MNRAS.373..799C}. Magnitude-independent noise sources such as scintillation and flat-fielding errors produce a noise floor $\sigma_s\simeq 0.004$ mag for the brightest targets. This performance can be approximated as 
\begin{equation}
\sigma_V\simeq\sqrt{\sigma_s^2+\frac{1}{f}},
\label{eq:sigma_v}
\end{equation}
where the effective photon count $f$ from a target of magnitude $V$ is $f \simeq 10^{4 -0.4(V-12)}$.

In the presence of correlated noise, the precision to which transit depth can be measured improves with the square root of the effective number of observations \citep{2006MNRAS.373..231P}. For low transit impact parameters, the duration of a transit is 
\begin{equation}
T\approx P\frac{R_s}{\pi a},
\end{equation} 
where $R_s$ is the stellar radius. The number of measurements within a single transit depends on the interval $\tau$ between visits, which for WASP is typically $\tau=600s$. The number of observations in a typical transit is thus $T/\tau$.

A typical ground-based observing season lasts $t_{\rm season}\simeq 120$ days. Allowing for a day/night duty cycle of 6 hours per day, the number of transits observed in a typical survey is $N_t\simeq t_{\rm season}/(4 P)$. In the absence of correlated noise, the  number of  in-transit  observations per season is approximately
\begin{equation}
N_{\rm eff}=N_{\rm tr}\frac{T}{\tau}=\frac{t_{\rm season} T}{4 \tau P}=\frac{t_{\rm season}}{4\tau}\frac{R_s}{\pi a}.
\end{equation}

\begin{figure}
\includegraphics[width=\columnwidth]{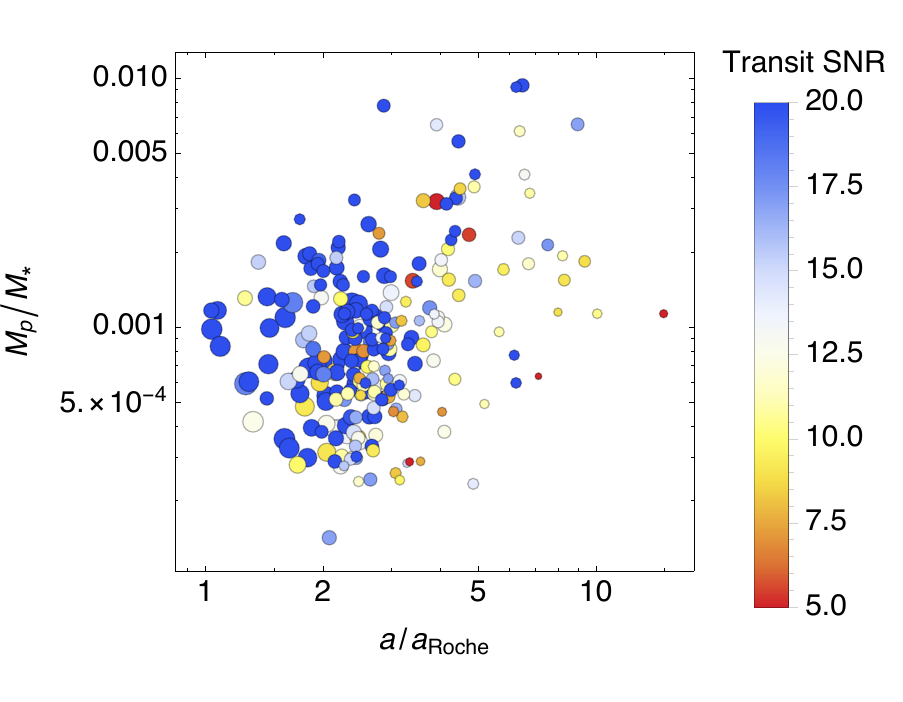}
\caption{Planet/star mass ratio as a function of orbital separation in units of the planet's Roche radius $a_{\rm Roche}$. Symbol size is proportional to planet radius, and symbol colour denotes the SNR of the transit detection, estimated using eqs.~\ref{eq:sigma_v} and \ref{eq:snr}.}
\label{fig:detection}
\end{figure}

The signal-to-noise ratio of a transit detection is then given by the ratio of the transit depth $\delta=(R_p/R_s)^2$ to the measurement precision,
\begin{equation}
{\rm SNR}=\frac{R_p^2}{R_s^2}\sqrt{\frac{t_{\rm season}}{4\tau \sigma^2_V}\frac{R_s}{\pi a }}.
\label{eq:snr}
\end{equation}
The mass-separation diagram is again presented in Figure~\ref{fig:mrocplot}. In this version, the colour coding of the symbols indicates that nearly all of the gas-giant planets from the ground-based surveys are detected at an estimated SNR$>12$ from the expression above. In Fig.~\ref{fig:detection} this detection threshold defines the lower right boundary of the scatter plot, separating strongly-detected blue-coloured points from more marginal yellow and red-hued points.

We caution that the SNR may not be the only contributor to this boundary. In scenarios where hot Jupiters are tidally circularised from highly eccentric orbits, one can obtain a curve of similar orientation and position by the requirement that the circularisation time of such a planet be under a reasonable limit. As noted in Section~\ref{sec:evap} above, however, eccentric migration and tidal circularisation ought to leave planets parked outside twice the Roche limit. This is clearly not the case for the full sample.

\section{Probability density model for a single system}
\label{sec:probdens}

Our goal is to deduce the posterior probability distribution for Qs' from the observed distribution of orbital separations $a$ of known transiting hot Jupiters from their host stars. 

\subsection{Definitions and assumptions}

In the context of this work, {\em parameters} are observationally-determined properties of the individual star-planet systems. The vector of relevant parameters for the $j$th star are the stellar effective temperature, mass and radius, and the planetary radius, mass and observed orbital period: $\boldsymbol{\theta}_j = \{T_{\rm eff}, M_s, R_s, M_p, R_p, P_{\rm obs}\}$. For this preliminary study we treat the parameters as fixed. In a full hierarchical Bayesian analysis we would marginalise over the posterior probability distributions of the parameters for the individual systems. In this largely analytic investigation, however, we assume that the system parameters are known to a precision such that the normalised probability density does not change significantly over their ranges. 

The {\em data} $\boldsymbol{x}$ are observables, whose values have statistical distributions that are sensitive to both the parameters of individual stars and the population-level quantities (hyperparameters) that we want to determine. In this study the data are the logarithmic scaled distances $x=\log(a/R_s)$ of individual planets from theirs star in units of the stellar radius.

The {\em hyperparameters} are global properties of the population, in this case the tidal quality factor $Q_s'$ and its dependence $\epsilon$ on individual stellar effective temperature. 

Together with the parameters of the individual stars and the nuisance parameters, the hyperparameters define the shape of the generative {\em model} distribution from which the data $\boldsymbol{x}$ are drawn. The form of the model is thus unique to each star-planet system. The rate of tidal orbit decay is inversely proportional to $Q_s'$, which has a parametrised dependence $\epsilon$ on stellar effective temperature. The model PDF $\rho(x_j|k(\boldsymbol{\theta}_j, Q_s', \epsilon), t_{\rm age,j})$ is the probability density as a function of star-planet separation $\log(a/R_s)$, given the hyperparameters and the parameters and age of the $j$th star. 

{\em Nuisance parameters} are parameters which affect the shape of the model PDF of the data, but whose values are either unknown or of no immediate interest. In this study, the stellar age is a nuisance parameter. The probability density at a given orbital separation is a function of system age, but we generally do not know the age of the star. We deal with it by marginalising over its (usually uninformative) prior distribution. That is, we replace the model of the PDF for the data at fixed age, with a weighted average 
\begin{equation}
\overline{\rho}(x_j|k(\boldsymbol{\theta}_j, Qs', \epsilon))=\int_0^{t_{\rm Life}}
\rho(x_j|k(\boldsymbol{\theta}_j, Qs', \epsilon), t_{\rm age,j})\pi(t_{\rm age,j})d t_{\rm age,j}
\end{equation}
of model PDFs integrated over the star's main-sequence lifetime $t_{\rm Life}$ using the normalised prior distribution $\pi(t_{\rm age})$ of the age as the weighting function. 

\subsection{Probability density model}

A planet of mass $M_p$ orbiting a star of mass $M_s$ and radius $R_s$ will evolve inward from its initial distance $a_{\rm init}$ throughout the star's nuclear-burning lifetime $t_{\rm nuc}$ under the influence of torques due to the tidal bulges raised on the host star (Eq.~\ref{eq:dlnadt}). 

An observer seeing a planet orbiting at scaled distance $r=a/R_s$ from the star is effectively sampling from an  ensemble of similar planets orbiting the same star, drawn from a statistical distribution of initial orbital separations. For a planet to be advected past the observed value of $r$, it must have been born at an initial distance $r_B$ from which it can pass $r$ during the star's nuclear-burning lifetime (or the age of the galactic disk if the star is sufficiently long-lived). 

The situation is analogous to the H-R diagram of a star cluster. Stellar masses are drawn from a continuous distribution, the initial mass function (IMF). The mass of a star defines its subsequent trajectory and speed along an evolutionary track in the H-R diagram. Evolutionary phases such as the Hertzsprung gap, where stars move rapidly across the H-R diagram,  are poorly represented in observational samples of finite size because stars evolve quickly through them. It's hard to catch a speeding bullet.

In the planetary tidal evolution problem, planets are born at a range of logarithmic birth distances $x_B=\log r_B$ from the star. This continuous initial distribution $\rho_0(x_B)$ can have any appropriate form, but for simplicity and approximate consistency with the giant-planet distribution from radial-velocity surveys \citep{2007ARA&A..45..397U}, we assume that systems are born with a  density distribution $\rho_0$ that is uniform with respect to $x_B$ for all planets irrespective of orbital inclination. This distribution is analogous to the stellar IMF, in that the birth distance defines the evolution of the orbital separation throughout the planet's remaining lifetime. 

The birth location of a system of age $t$ observed at present location $r$ is found by rearranging and integrating Eq.~\ref{eq:xrate} to obtain
\begin{eqnarray}
\int_0^t d\tau&=&-\frac{1}{k}\int^{r}_{r_B}r^{11/2}dr\nonumber\\
\Rightarrow r_B^{13/2}&=&r^{13/2} +\frac{13}{2}k t.
\label{eq:dr}
\end{eqnarray}
In a snapshot survey, the probability density of observing a planet at its present orbital separation $r$ depends on both the initial distribution of orbital separations and the planet's rate of inward drift at position $r$. The migration rate determines the range of $r$ within which we might catch the planet at the time of the survey, and is thus inversely proportional to the probability density at any position. We have assumed that $\rho_0$ is distributed uniformly in $x_B=\log r_B$, so it is convenient to use logarithmic distances $x=\log r$. By analogy with steady one-dimensional fluid flow, the flux $\rho(x)|dx/dt|$ is a constant. The probability density $\rho$ of catching a planet at the observed position $x$ thus depends on the density $\rho_0(x_B)$ at $x_B$ and the ratio of the inward migration rates $|dx/dt|$ at $x$ and $x_B$. 

An additional geometric factor $e^{-x}=a/R_s$ is needed to account for the decrease in transit probability with distance from the star. The density at the observed location $x$ is then
\begin{eqnarray}
\rho(x|k,t)&=&\rho_0 \left.\frac{dx}{dt}\right|_{x_B}\left(\left.\frac{dx}{dt}\right|_x\right)^{-1}e^{-x}=\rho_0\frac{e^{-13x_B/2}}{e^{-13x /2}}e^{-x}\nonumber\\
&=&\rho_0\frac{e^{13x/2}e^{-x}}{e^{13x /2}+13 k t / 2}.
\label{eq:rhoxt}
\end{eqnarray}

\subsection{Marginalising over stellar age}

Since the age $t$ of the star is generally unknown, or at best constrained by a prior probability density distribution $\pi(t)$, we need to marginalise the density at $x$ over the range of possible ages $0 < t < t_{\rm Life}$. Here the stellar lifetime is the lesser of the main-sequence nuclear lifetime and the age of the galactic population to which the star belongs, $t_{\rm Life}={\rm Min}(t_{\rm nuc},t_{\rm gal})$:
\begin{equation}
\overline{\rho}(x|k)=\int_0^{t_{\rm Life}}\rho(x|k,t)\pi(t)dt.
\end{equation}

\begin{figure}
\includegraphics[width=\columnwidth]{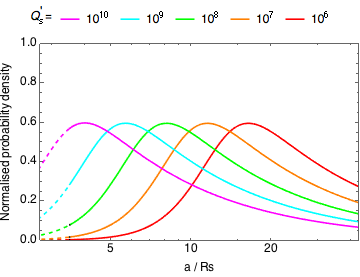}
\includegraphics[width=\columnwidth]{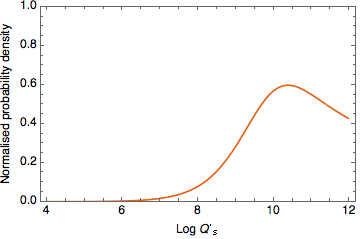}
\caption{The upper panel shows the age-marginalised PDF for hypothetical planets orbiting WASP-18. Smaller values of the tidal quality factor $Q_s'$ result in faster migration, depleting the inner regions of planets. The boundary between the solid and dotted curves is at the location of WASP-18b. The probability density at this position is plotted as a function of $Q_s'$ in the lower panel. WASP-18b is a massive planet orbiting very close to its host star, so its existence favours large values of $Q_s'$. The dotted lines extend all the way to the Roche limit at 1.19 $R_s$; the planet is too massive to undergo evaporation close to the star.}
\label{fig:wasp18PDF}
\end{figure}

\begin{figure}
\includegraphics[width=\columnwidth]{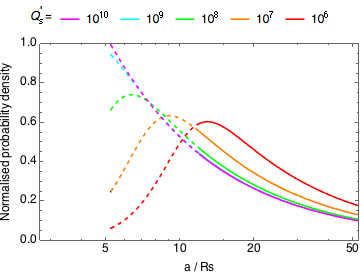}
\includegraphics[width=\columnwidth]{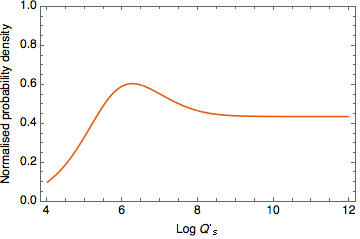}
\caption{The upper panel shows the age-marginalised PDF for hypothetical planets orbiting WASP-21. The boundary between the solid and dotted curves is at the location of WASP-21b. The probability density at this position is plotted as a function of $Q_s'$ in the lower panel. WASP-21b is a low-mass planet planet orbiting relatively far from its host star, so its existence is consistent with a wide range of $Q_s'$. The inner limit for the normalisation integral lies at 4.86 $R_s$, the evaporation limit for this planet. The Roche limit is further in, at 3.87 $R_s$.}
\label{fig:wasp21PDF}
\end{figure}

A uniform prior $\pi(t)=\mathcal{U}(0,t_{\rm Life})$  gives
\begin{equation}
\overline{\rho}(x|k)=\frac{\rho_0}{t_{\rm Life}}\int_0^{t_{\rm Life}}\frac{e^{13x/2}e^{-x}}{e^{13 x/2}+13 k t / 2}dt
\end{equation}
which has the analytic solution
\begin{equation}
\overline{\rho}(x|k)=\rho_0\frac{\log(1+(13/2)kt_{\rm Life}e^{-13x/2})}{(13/2)kt_{\rm Life}e^{-13x/2}}e^{-x}.
\label{eq:rhox}
\end{equation}
For $\overline{\rho}(x|k)$ to be used as a generative model for calculating likelihoods, it must be normalised by integrating from the smallest possible separation $x_{\rm min}={\rm Max}(1,x_{\rm Roche},x_{\rm evap})$ to a separation large enough to capture most of the probability. The integral 
\begin{equation}
\int_a^b \frac{\log(1+Ce^{-13x/2})}{Ce^{-13x/2}}e^{-x}dx
\end{equation}
has an analytic, but rather lengthy solution that is easily found and evaluated with a symbolic algebra package. 

The families of  age-marginalised PDFs for hypothetical transiting planets orbiting the stars WASP-18 and WASP-21 are illustrated
in Figures~\ref{fig:wasp18PDF} and \ref{fig:wasp21PDF}.

\subsection{Bayesian determination of $Q_s'$}

If we assume that all stars have comparable values of $Q_s'$, the likelihood of obtaining the set of orbital distances in the sample is the product of all the generative model probabilities described above for the targets' orbital separations:
\begin{equation}
\mathcal{L}(\boldsymbol{x}|Q_s',\epsilon)=\prod_{j=1}^N \overline{p}(x_j|k(\boldsymbol{\theta_j},Q_s',\epsilon)),
\label{eq:likelihood}
\end{equation}
where the index $j$ refers to individual stars with parameters $\boldsymbol{\theta}_j=\{M_s,R_s,M_p,R_p,T_{\rm eff}\}_j$. Here we use the symbol $\overline{p}(x|k)$ to denote the normalised version of $\overline{\rho}(x|k)$. The model is conditioned on the parameters and hyperparameters through the coefficient $k(\boldsymbol{\theta_j},Q_s',\epsilon)$. The likelihood is shown as a function of $Q_s'$ in the lower panels of Figs.~\ref{fig:wasp18PDF} and \ref{fig:wasp21PDF}. 

Before attempting to infer $Q_s'$ from the observed sample of transiting hot Jupiters, however, we set about testing the method on mock datasets with properties designed to be as similar as possible to the observed sample.

\section{Mock catalogues}
\label{sec:mockcat}

In order to determine whether population-level parameters can be extracted from the measured physical parameters of the known transiting hot-Jupiter systems, we created mock catalogues in which the parameters controlling selection effects and orbital migration were tunable. The initial stellar sample was drawn from  Version 1.6 of the TRILEGAL galaxy model \citep{2012ASSP...26..165G}. A field of 10 deg$^2$ in the direction of the galactic pole was sampled in the magnitude range $8.5<R<12.0$. The selection emulates the saturation and detection limits for hot-Jupiter transits in the WASP and HAT surveys, which have produced the majority of the known transiting hot Jupiters. Every star in the sample is characterised by its mass $M_s$, radius $R_s$, effective temperature $T_{\rm eff}$ and age $t_s$.

\subsection{Planet masses and radii}
Each star was allocated a planet with a mass drawn from a lognormal distribution with a mean $\log_{10}M_p=0.046$ and $\sigma(\log_{10}M_p)=0.315$, derived from the sample of well-studied transiting planets in the TEPCat catalogue of \cite{2011MNRAS.417.2166S} with masses in the range $0.3 < M_p < 13.0$ M$_{\rm Jup}$. Initial planetary orbital periods $P_{\rm init}$ were sampled from a uniform distribution in log period. The upper bound of the sampling range is set at an arbitrary $P_{\rm max}=40$ days. The lower bound of the sampling range corresponds to the Roche limit
\begin{equation}
a_R=2.46 R_p\left(\frac{Ms}{Mp}\right)^{1/3}.
\end{equation}


\begin{figure}
	\includegraphics[width=\columnwidth]{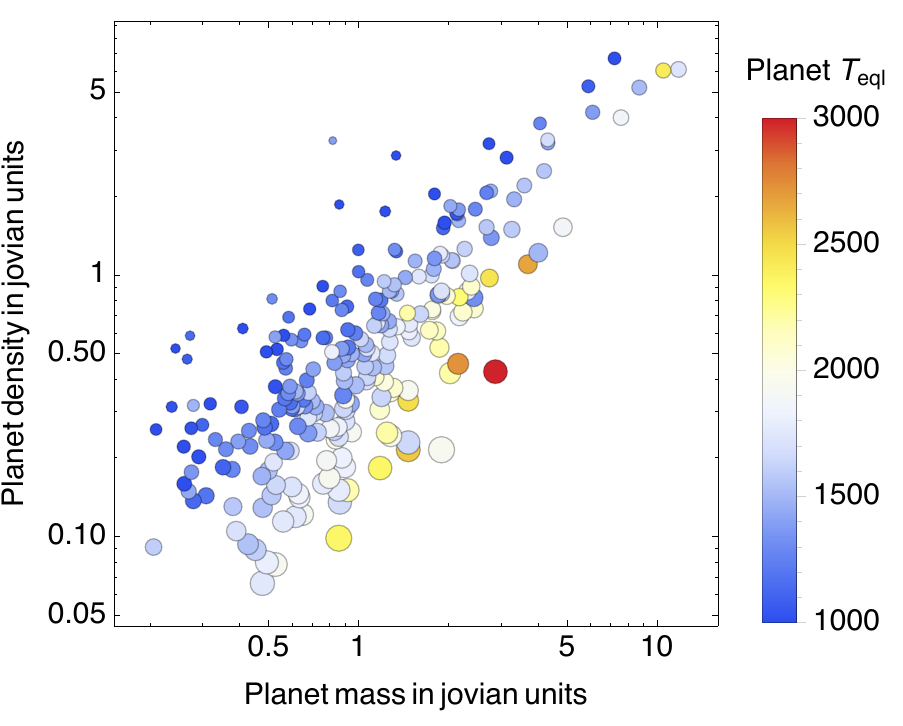}
    \caption{Bulk densities of transiting exoplanets in TEPCat as a function of planetary mass. Symbol size is directly proportional to planet radius. Symbol colour denotes planetary blackbody equilibrium temperature. Cold planets have radii nearly independent of mass, so density scales linearly with mass. Strongly-irradiated planets have consistently larger radii and hence lower densities than their cooler counterparts of the same mass, particularly at the lower end of the mass range.}
    \label{fig:mrhoplot}
\end{figure}

\begin{figure*}
\includegraphics[width=\columnwidth]{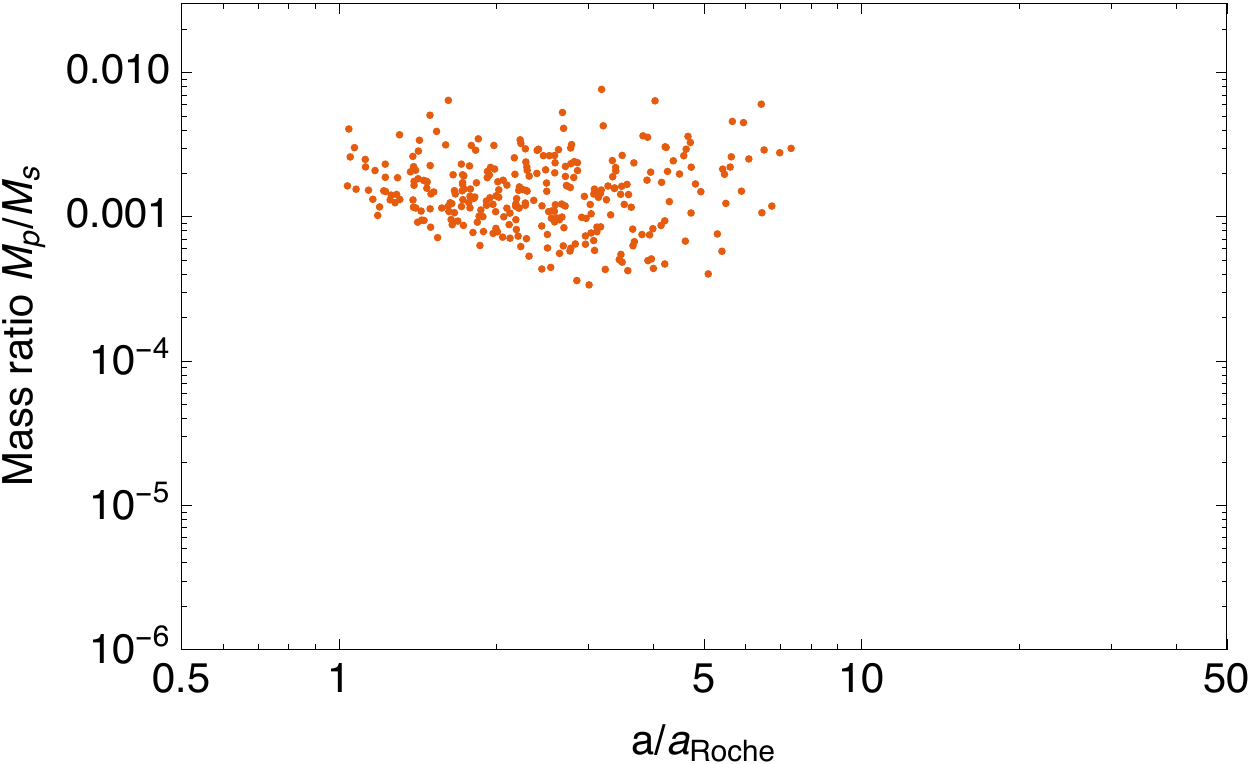}
\includegraphics[width=\columnwidth]{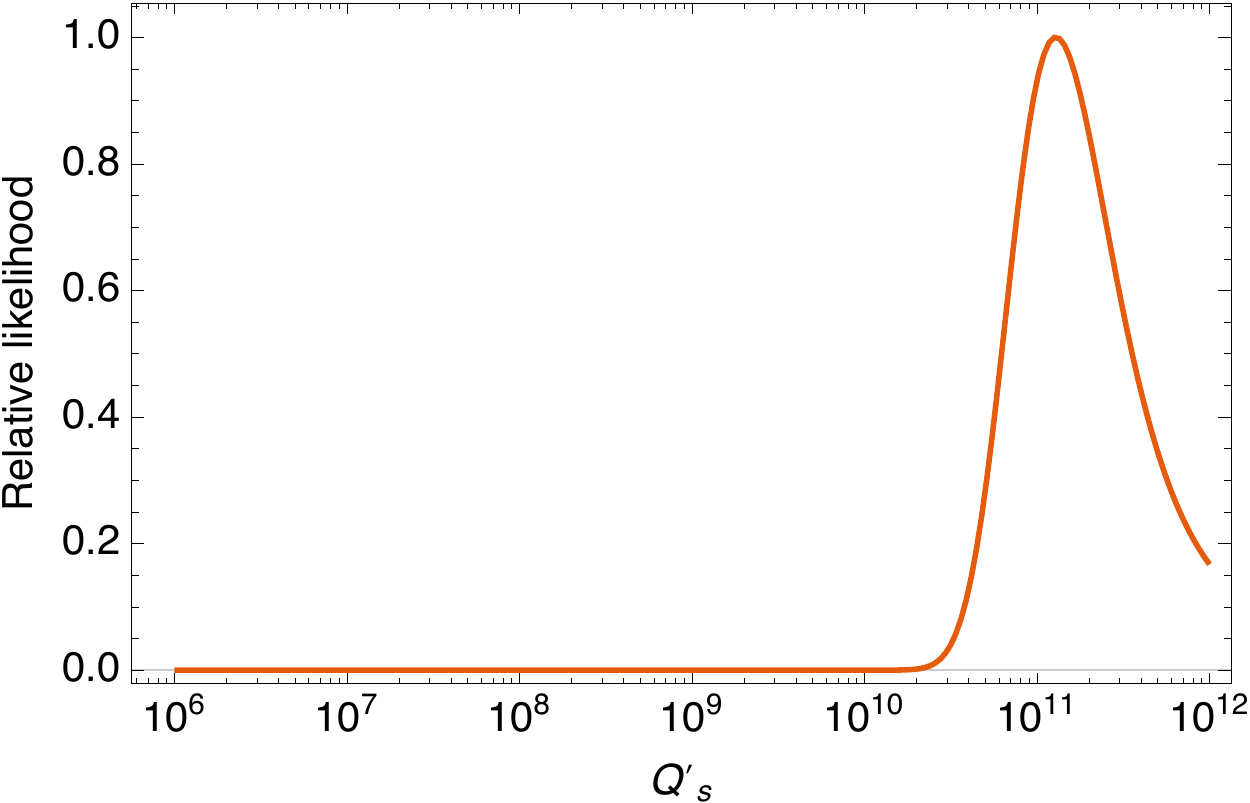}
\includegraphics[width=\columnwidth]{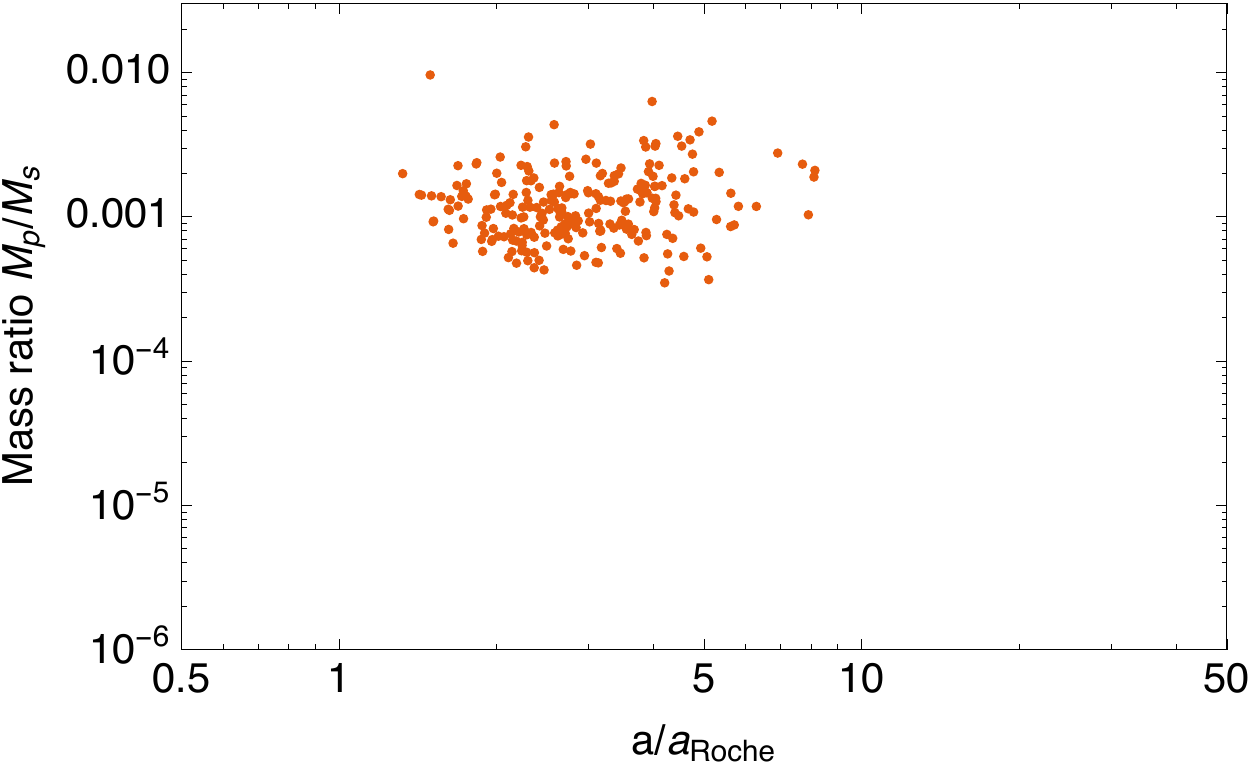}
\includegraphics[width=\columnwidth]{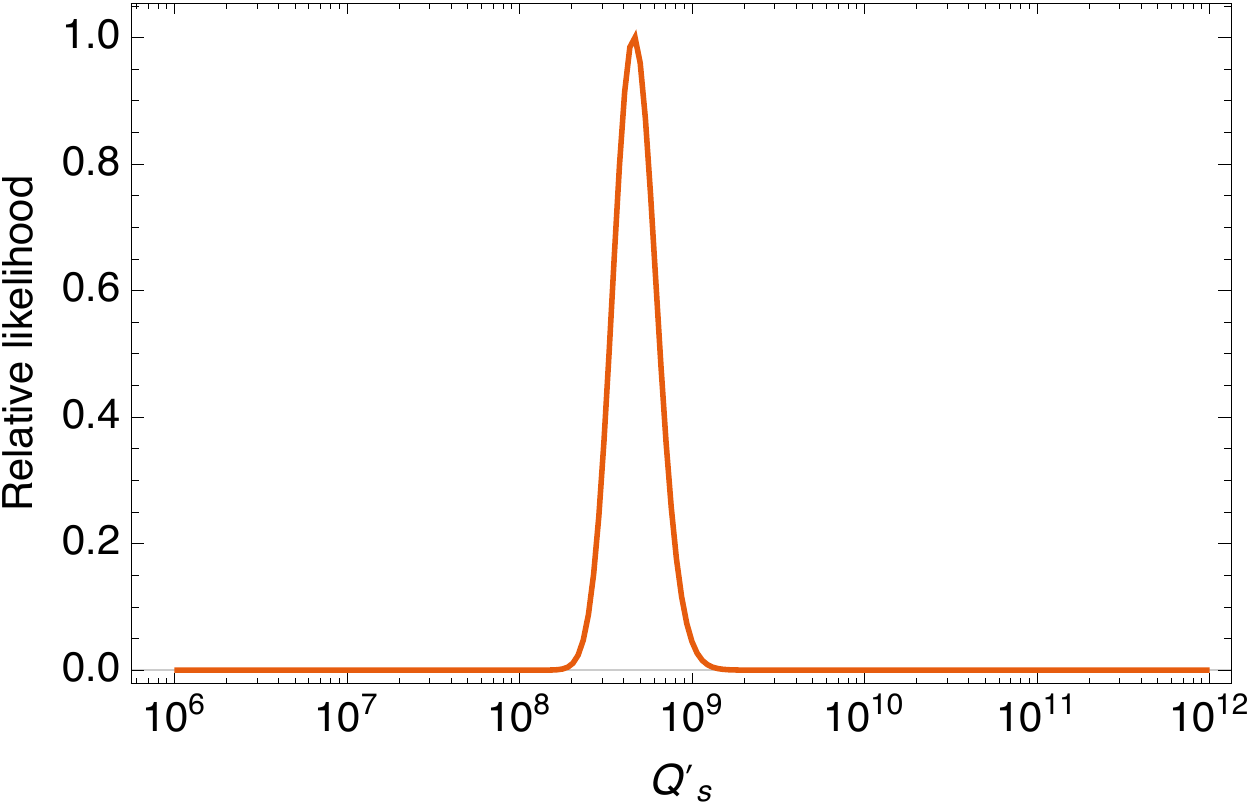}
\includegraphics[width=\columnwidth]{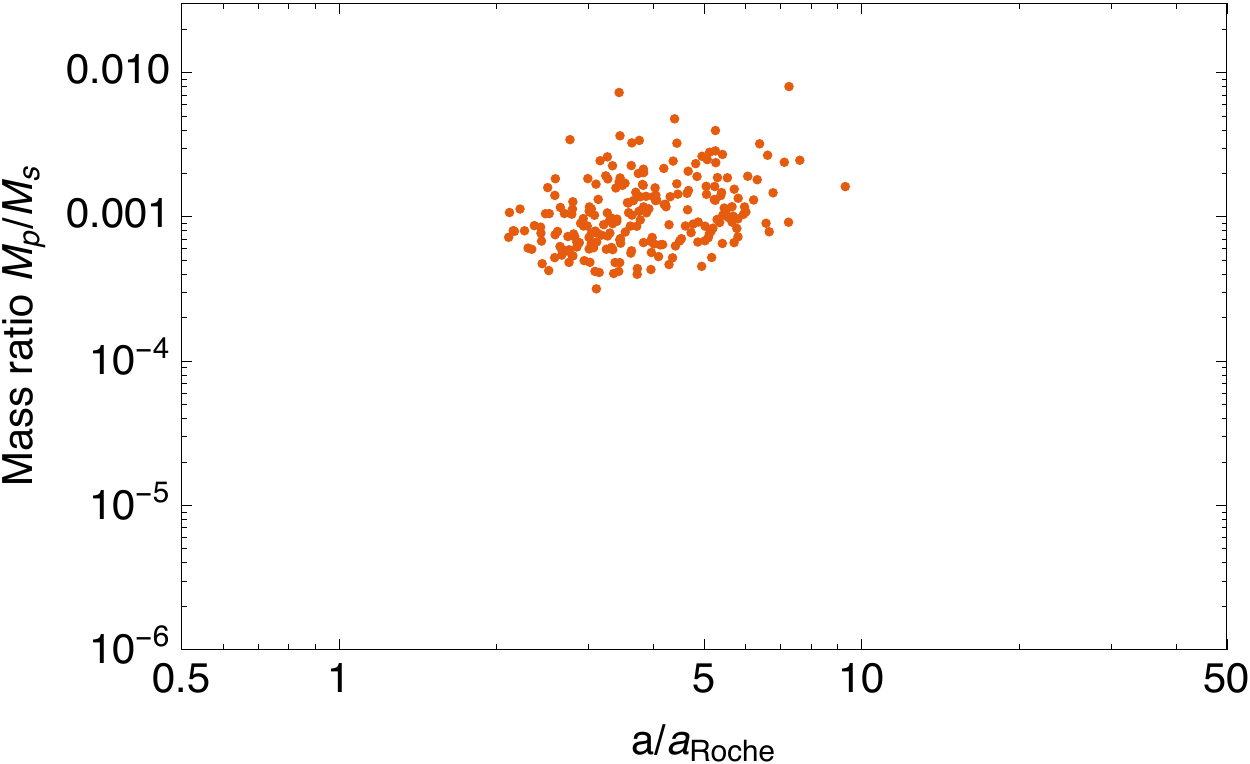}
\includegraphics[width=\columnwidth]{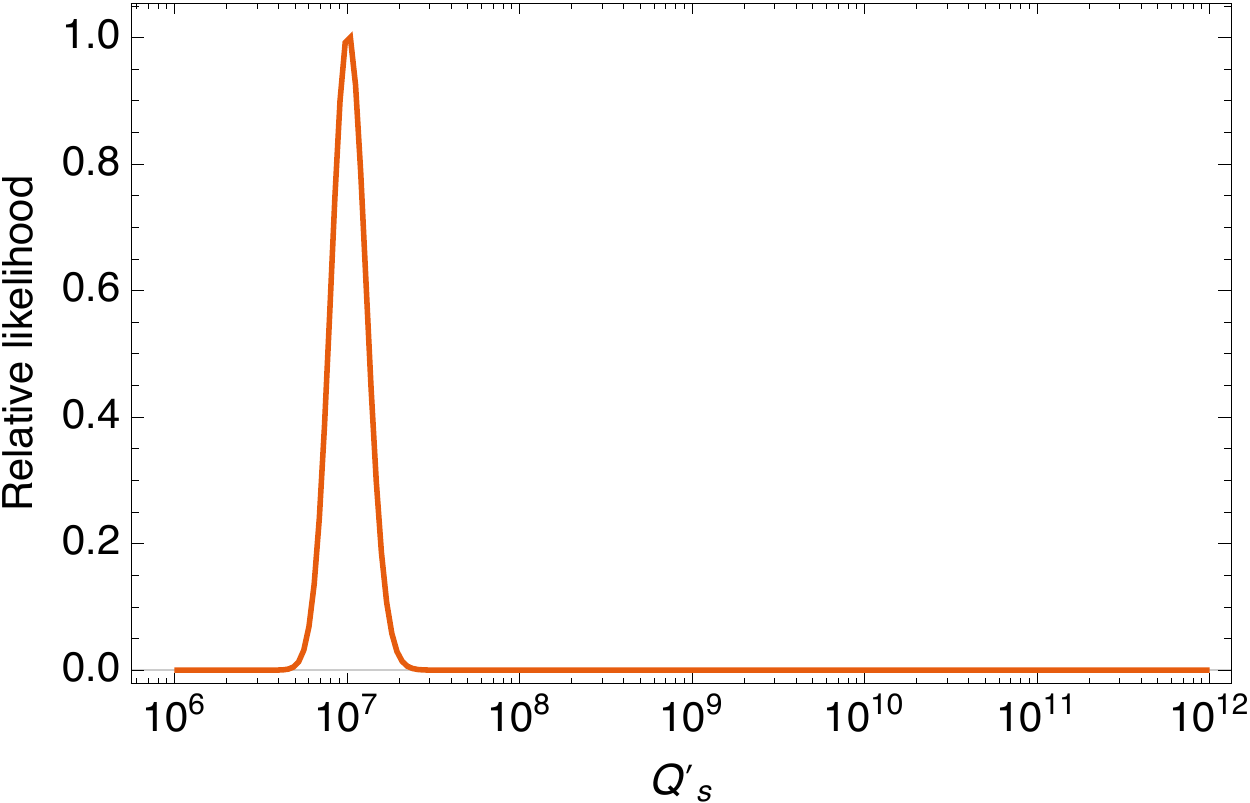}
\caption{Mass-separation diagrams and relative likelihood as a function of $Q_s'$ for three of six mock catalogues of transiting planets, generated with tidal quality factors  $Q_s'=10^{11}$ (top),  $10^{9}$(middle) and $10^{7}$ (bottom).}
\label{fig:mockmassep}
\end{figure*}

The planet radius $R_p$ is known to be sensitive to irradiation. Figure~\ref{fig:mrhoplot} shows the almost-linear dependence of density on mass for cold planets, modified by the envelope expansion that occurs in the presence of strong irradiation. Modifying the approach of \cite{2010A&A...516A..33E} for the present sample, we fitted a cubic dependence of log planetary radius on log mass and a linear dependence on log equilibrium temperature, obtaining:
\begin{eqnarray}
\log_{10} \left(\frac{R_p}{R_{\rm Jup}}\right)&=&
c_0
+c_1\log_{10}\left(\frac{M_p}{0.94 M_{\rm Jup}}\right)\nonumber\\
&&+c_2\log_{10}\left(\frac{M_p}{0.94 M_{\rm Jup}}\right)\nonumber\\
&&+c_3\log_{10}\left(\frac{M_p}{0.94 M_{\rm Jup}}\right)\nonumber\\
&&+c_4\log_{10}\left(\frac{T_{\rm eql}}{1471\ {\rm K}}\right)\label{eq:radfit},
\end{eqnarray}
where $c_0=0.1195$, $C_1=-0.0577$, $c_2=-0.1954$, $c_3=0.1188$, and $c_4=0.5223$. 
The resulting radius $R_p$ was assigned to each synthetic planet according to its mass and blackbody equilibrium temperature $T_{\rm eql}=T_{\rm eff}(Rs/2a)^{1/2}$, computed assuming zero albedo and isotropic re-radiation. 

\subsection{Observational selection}

We applied simple rejection criteria to simulate the planet-detection capabilities of the ground-based wide-field transit surveys. We sampled from a random distribution of orbital inclinations, selecting only those systems which would be transiting according to the sampling criterion $\mathcal{U}(0,1)<R_s/a$, where $\mathcal{U}(0,1)$ is a single sample drawn from a uniform distribution between 0 and 1.
Systems with ${\rm SNR}<12$ (cf. Fig.~\ref{fig:detection}) were rejected as yielding transits too shallow or too noisy for reliable detection, effectively removing all giants, upper-main sequence stars and undetectably small planets from the sample.

\subsection{Tidal and evaporative erosion of mock planets}

The present-day orbital separation $a$ of every detectable planet was determined by integrating Eq.~\ref{eq:dlnadt} from the value $a_{\rm init}$ corresponding to $P_{\rm init}$ at $t=0$, forward in time to $t=t_s$. Planets which had decayed to orbital separations $a<{\rm Max}(a_R,R_s,a_{\rm evap})$ by the star's present age were removed.

\section{Parameter retrieval tests}

\subsection{Tidal quality factor retrieval}
\label{sec:retrievaltests}

We created six mock catalogues according to the recipe in Section~\ref{sec:mockcat}, with $Q_s'=10^6$, $10^7$, $10^8$, $10^9$ and $10^{10}$ and $10^{11}$.  In each case we set a detection threshold ${\rm SNR}=12$ and applied all the rejection criteria. We kept generating new systems until the size of the mock catalogue was comparable to the 236 stars included in the TEPCat sample. The resulting mass-separation diagrams are shown in Fig.~\ref{fig:mockmassep}. Note the progressive erosion of massive, short-period planets as the tidal quality factor is decreased and the rate of tidal orbit decay increases.

For every star $j$ in each of the six catalogues we computed $\overline{p}(x_j|k(\boldsymbol{\theta}_j,Q_s',\epsilon))$ over the domain $10^6<Q_s'<10^{12}$. The likelihood was then computed using eq.~\ref{eq:likelihood}. The likelihood is plotted as a function of $\log_{10}Q_s'$ in the right-hand panels of Fig.~\ref{fig:mockmassep}. The recovered estimates of $\log_{10}Q_s'$ are very close to the values used to generate the catalogues for $10^6<Q_s'<10^{11}$ (see also Table~\ref{tab:mockvalues}).

\begin{figure*}
\includegraphics[width=\columnwidth]{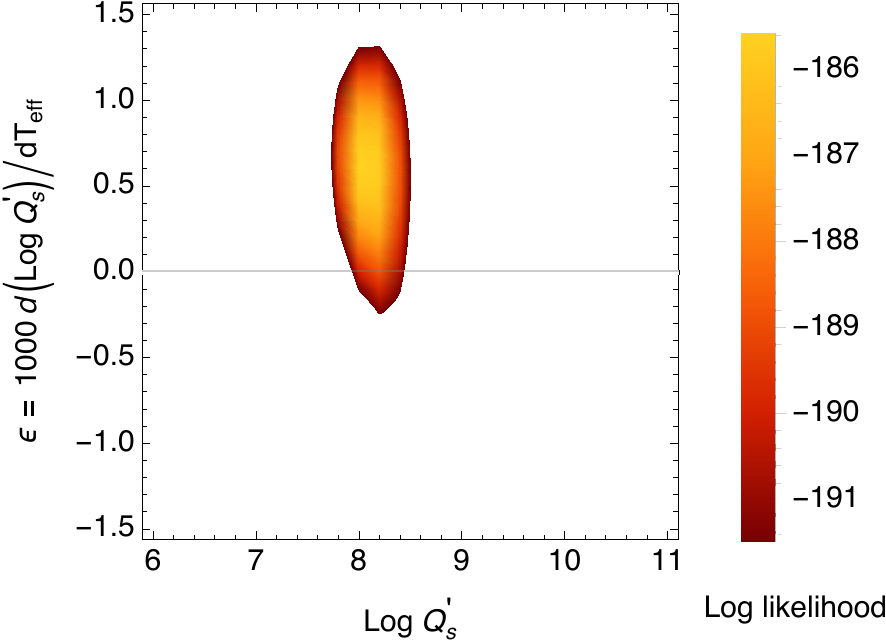}
\includegraphics[width=\columnwidth]{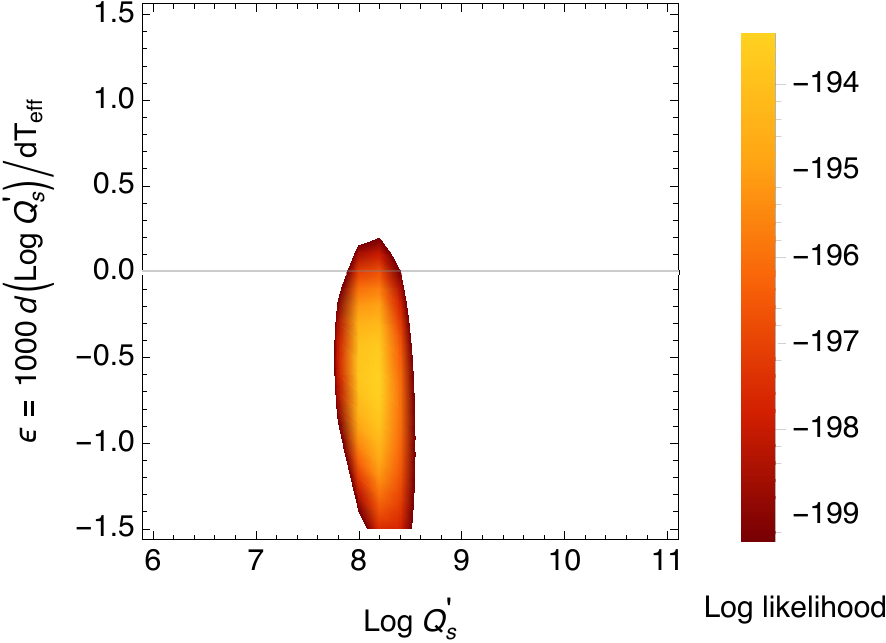}
\caption{Log likelihood as a function of $\log_{10} Q_s'(5850\ {\rm K})$ and $ \epsilon$ for two mock catalogues of transiting planets, generated with tidal quality factor $\log_{10} Q_s'(5850\ {\rm K})=8.3$ and $\epsilon = +0.7$ (left)
and $\epsilon = -0.7$ (right). In each case the range of the colour table covers a range $\Delta (2\log\mathcal{L}) =11.8$, corresponding to the $3-\sigma$ confidence region for a 2-parameter fit.}
\label{fig:mocklikely2D}
\end{figure*}

The good agreement between the input and recovered values of $\log Q_s'$ confirms that parameter retrieval works successfully for mock samples generated with a uniform initial distribution in log period, which is broadly consistent with expectations from Type II disc migration. The probability density function is modelled using the same assumption.  In reality, however, the initial distribution may not be uniform in $\log P_{\rm init}$. Real populations of hot Jupiters show a mix of orbital obliquities and eccentricities \citep{2010A&A...524A..25T,2012ApJ...757...18A}. This suggests that two or more distinct migration mechanisms may operate, giving rise to sub-populations with different initial period distributions \citep{2015ApJ...798...66D}. For example, \cite{2006ApJ...638L..45F} modelled the orbital separation distribution that would result from eccentric migration followed by tidal circularisation. They found that it produced a roughly gaussian distribution in $a_{\rm init}/a_{\rm Roche}$, with a mean orbital separation close to 3 $a_{\rm Roche}$. We tested the robustness of the retrieval procedure to the form of the initial period distribution by carrying out a second set of simulations, drawing each planet's $\log P_{\rm init}$ from a normal distribution centred on a value corresponding to $a_{\rm init}=3 a_{\rm Roche}$, and a standard deviation of 0.2 in the natural log. The systems in the mock sample were then evolved in the same way as described above, and the tidal quality factors retrieved. 

The full results, listed in Table~\ref{tab:mockvalues}, verify that the value of $Q_s'$ can be recovered reliably (within a factor 2) from mock samples of transiting hot Jupiters subject to the astrophysical and observational selection effects that define the sample of such planets found by ground-based, wide-field photometric surveys. The fidelity of the retrieval is excellent over the full range $6 < \log Q_s' < 11$ for mock samples with uniform distributions of $\log P_{\rm init}$. For populations with a narrow gaussian distribution in $\log P_{\rm init}$, piled up just outside twice the Roche limit, the retrieved values show a good match to the input values over the range $6.5 <  \log Q_s' < 8.5$. For $\log Q_s' > 8.5$, however, the retrieved value of the tidal quality factor saturates at $\log Q_s'\simeq 8.7$. We conclude that our method is robust even for planet populations whose initial period distributions resulted from a variety of migration paths.

\begin{table}
\caption{Results of parameter retrieval tests derived from mock catalogues. 
The first column gives the input values of $Q_s'$.
The second and fourth columns give the number of systems generated in each simulation.
The third and fifth columns give the inferred values of $\log Q_s'$ for
uniform and gaussian distributions of $\log P_{\rm init}$, as described 
in the main text.}
\label{tab:mockvalues}
\begin{tabular}{ccccc}
$\log Q_s'$ & $N_{\rm sys}$ & $\log Q_s'$ & $N_{\rm sys}$ & $\log Q_s'$ \\
            &               & (Uniform $\log P_{\rm init}$) && (Gaussian $\log P_{\rm init}$) \\
\hline\\
11.0 & 293 & $11.10\pm 0.30$ & 266 & $8.91 \pm 0.14$ \\
10.0 & 287 & $9.77 \pm 0.16$ & 264 & $8.65 \pm 0.13$ \\
9.0  & 243 & $8.66 \pm 0.13$ & 228 & $8.43 \pm 0.12$ \\
8.0  & 206 & $7.71 \pm 0.12$ & 245 & $7.93 \pm 0.10$ \\
7.0  & 229 & $7.01 \pm 0.10$ & 233 & $7.33 \pm 0.10$ \\
6.0  & 239 & $6.14 \pm 0.10$ & 228 & $6.55 \pm 0.10$ \\
\hline\\
\end{tabular}
\end{table}

\subsection{Dependence of $Q_s'$ on stellar $T_{\rm eff}$}

\begin{figure*}
\includegraphics[width=\columnwidth]{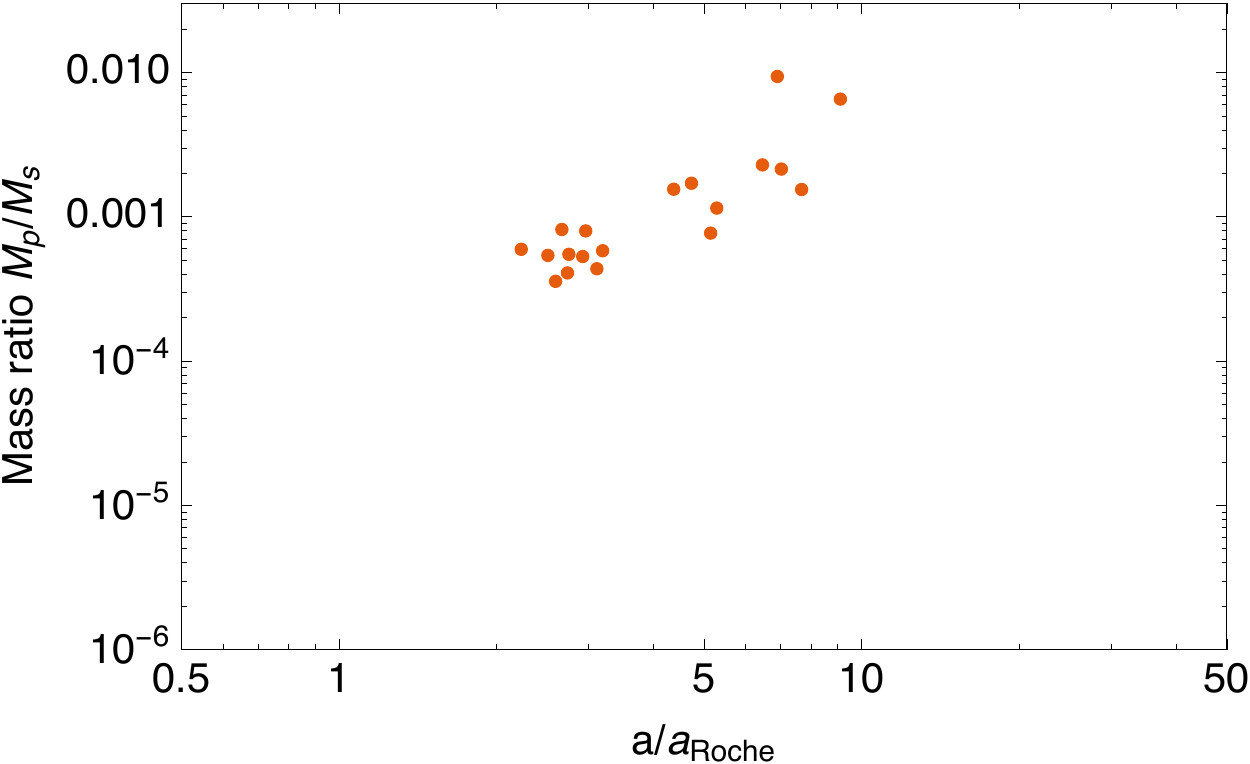}
\includegraphics[width=\columnwidth]{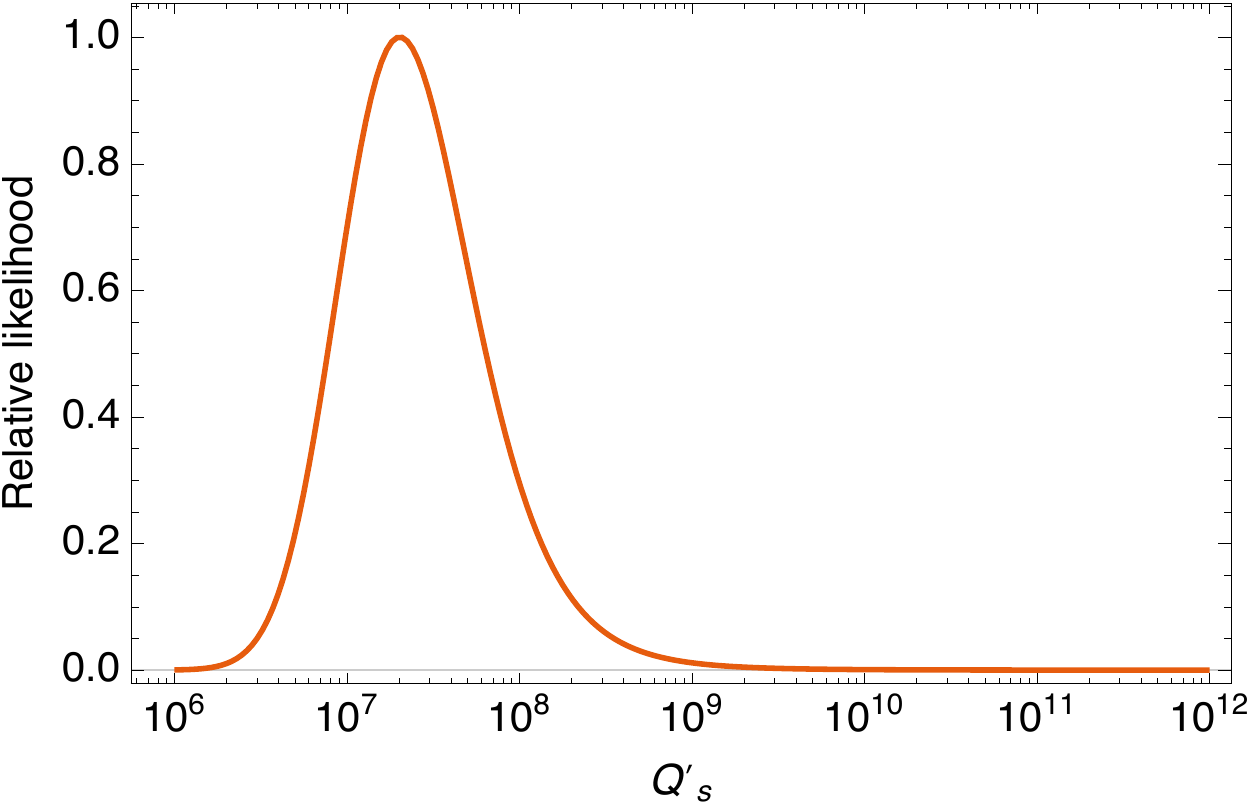}
\includegraphics[width=\columnwidth]{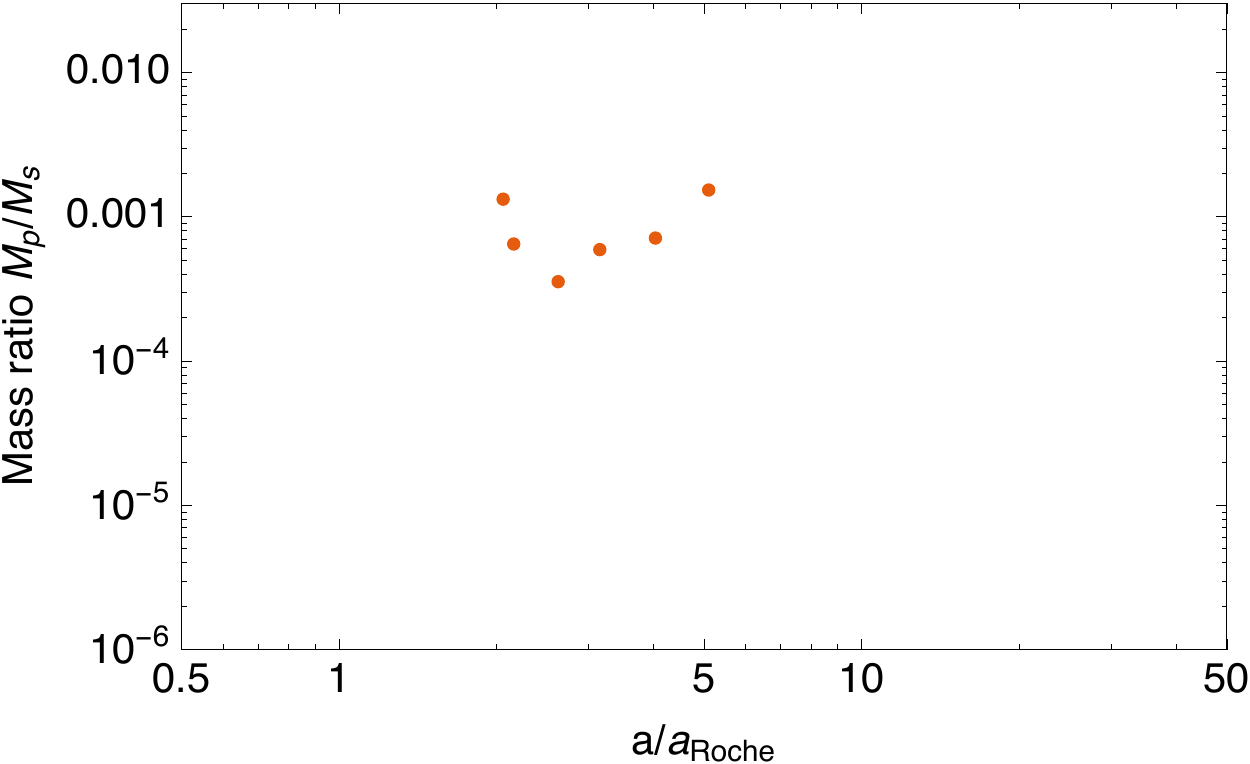}
\includegraphics[width=\columnwidth]{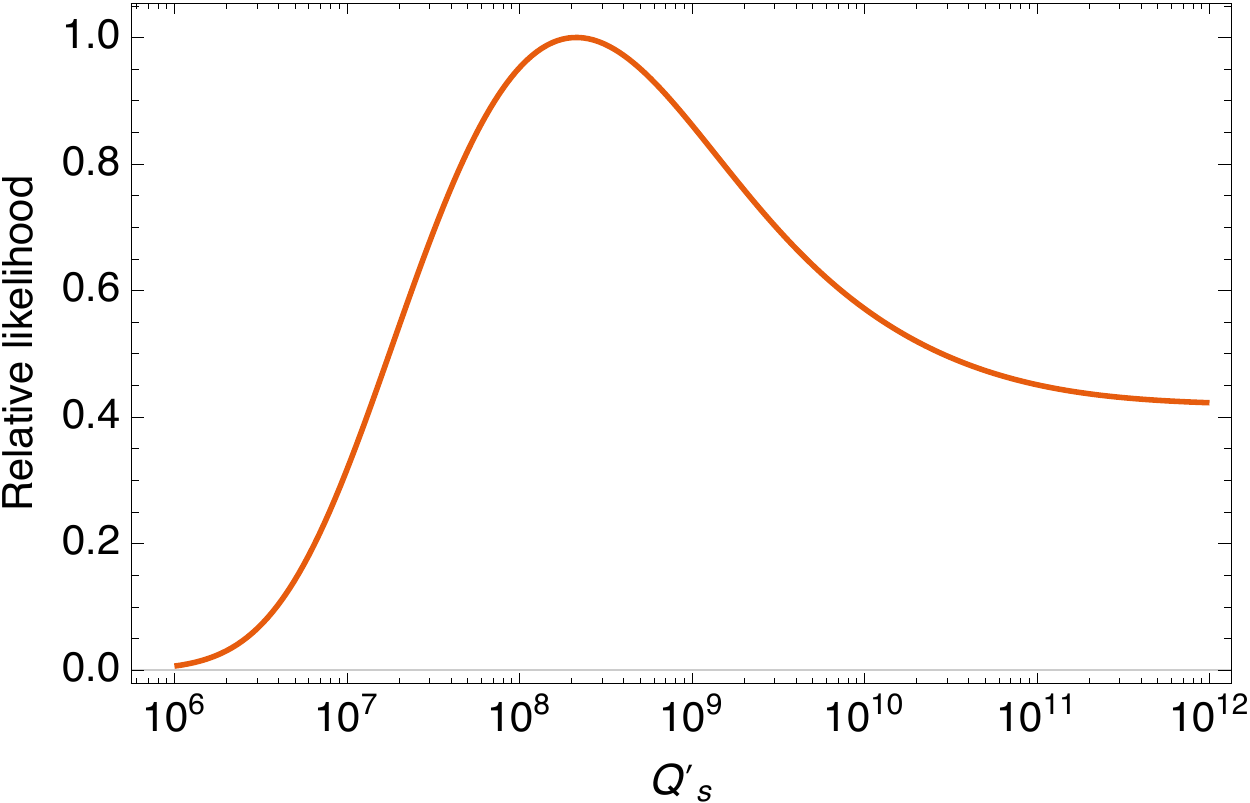}
\includegraphics[width=\columnwidth]{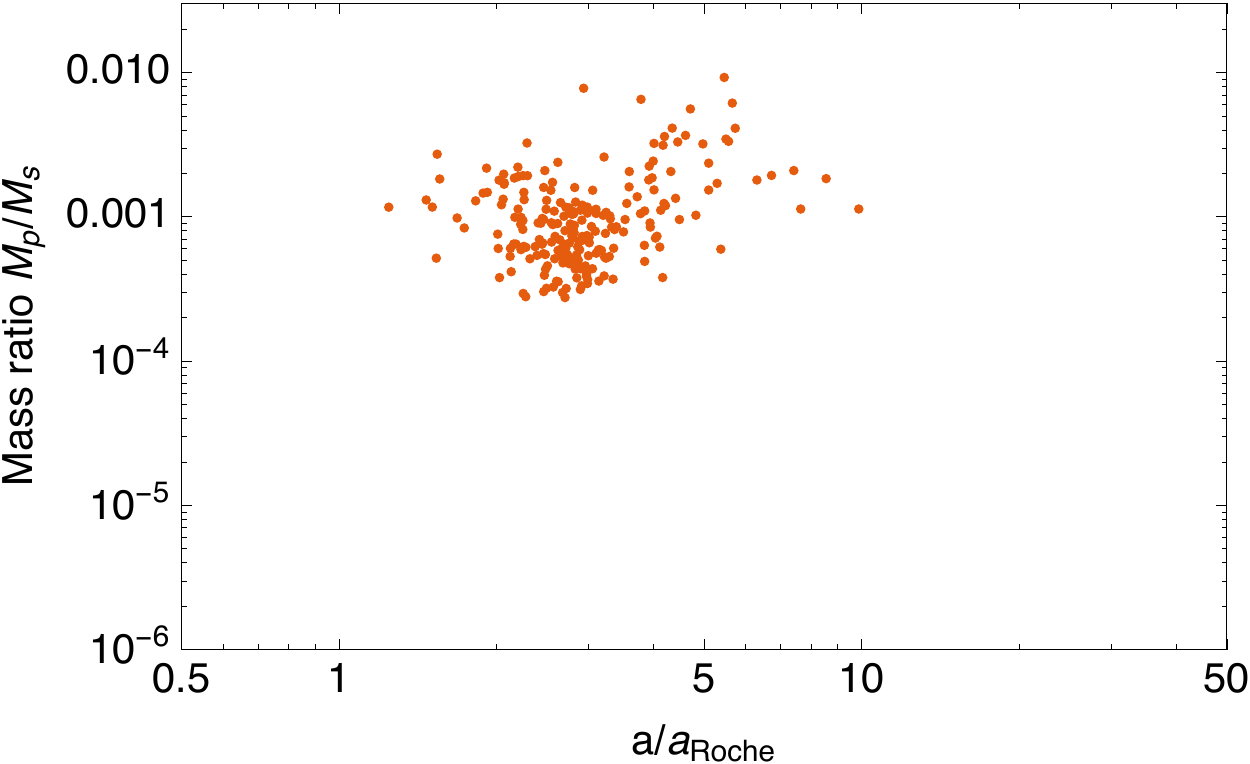}
\includegraphics[width=\columnwidth]{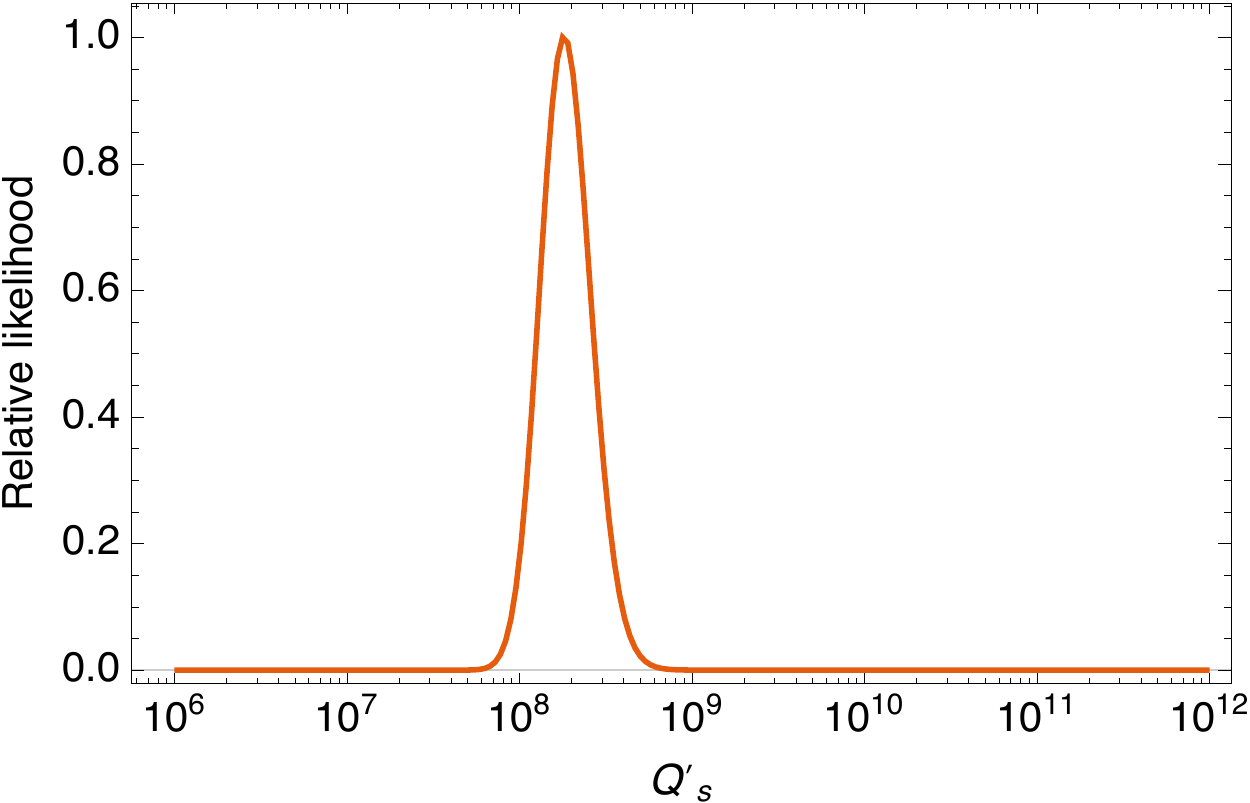}
\caption{Mass-separation diagrams and relative likelihood as a function of $Q_s'$ for the three sub-samples of transiting hot Jupiters extracted from TEPCat: 19 prograde planets with $0.5<P_{\rm orb}/P_{\rm rot}<2.0$; six highly-inclined or retrograde  planets in the same period-ratio range; and the full sample with the 19 prograde planets excluded.}
\label{fig:realdata}
\end{figure*}

\begin{figure}
\includegraphics[width=\columnwidth]{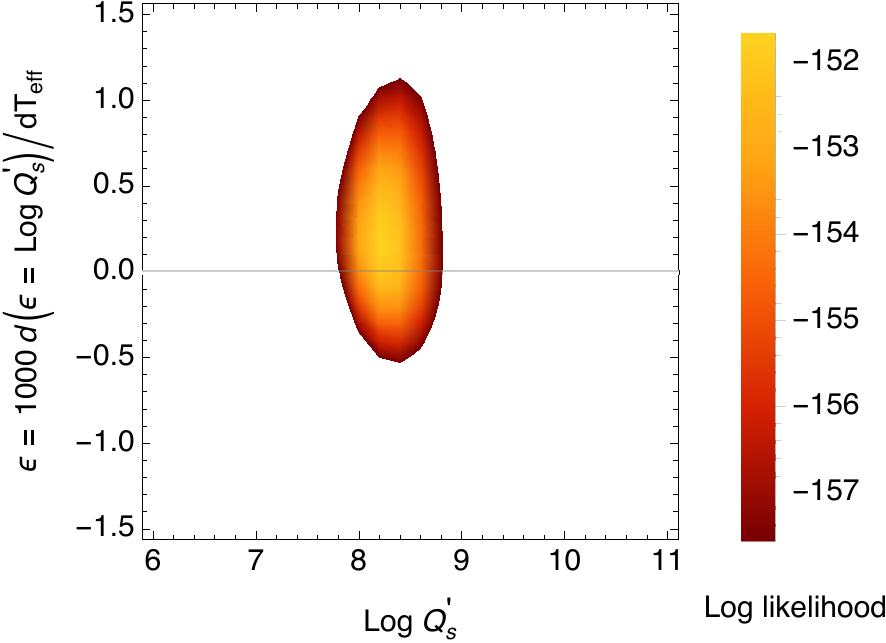}
\includegraphics[width=\columnwidth]{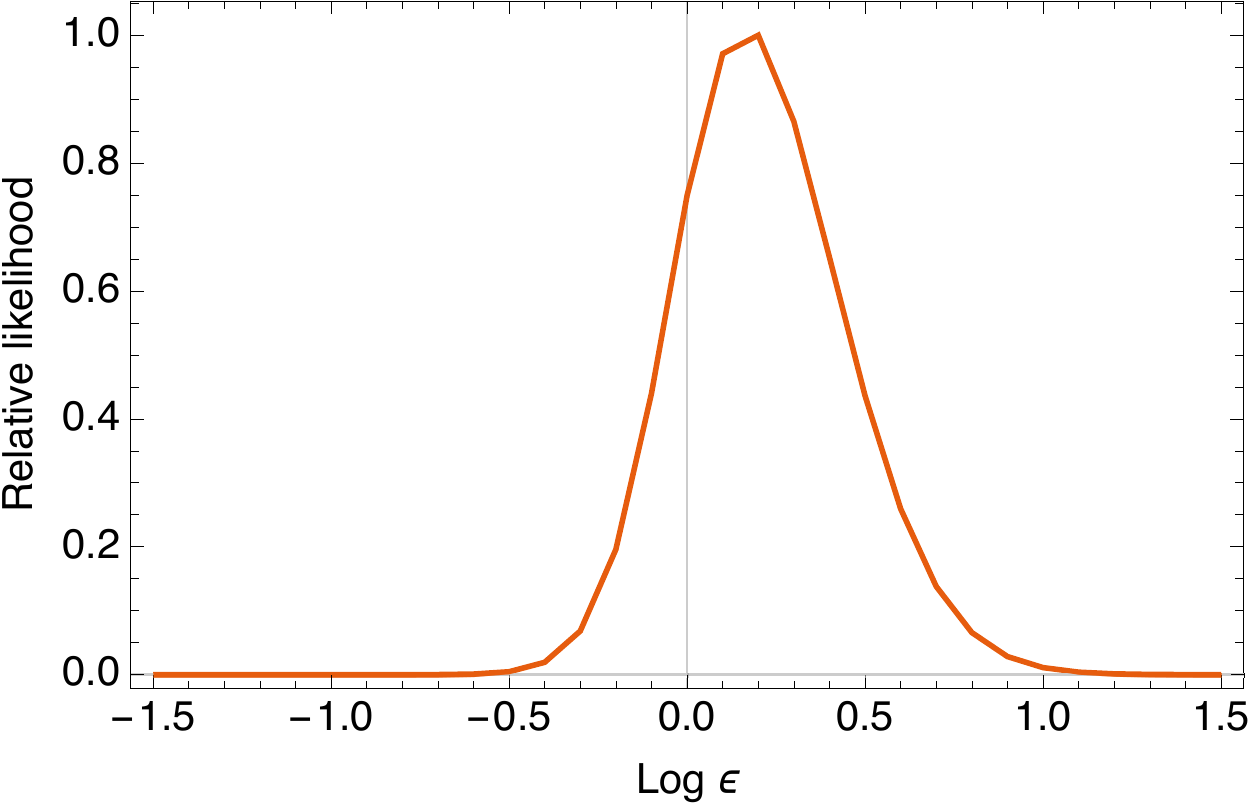}
\caption{The log likelihood map for the temperature-dependent model of the tidal quality factor, for the sample of transiting hot Jupiters extracted from TEPCat, with 19 prograde planets having $0.5<P_{\rm orb}/P_{\rm rot}<2.0$ excluded. The lower panel shows the relative likelihood as a function of the coefficient $\epsilon$ of temperature dependence, showing no significant evidence that $Q_s'$ increases with stellar effective temperature.}
\label{fig:likely2D}
\end{figure}

So far we have assumed that a single value of $Q_s'$ is applicable to the entire set of host stars. There are many good physical reasons why this might not be the case. \cite{2011ApJ...731...67P} find that the rate of turbulent convective dissipation decreases between models with masses of 0.8 M$_\odot$ and 1.4 M$_\odot$, suggesting an increase in  the coupling strength with increasing convective-zone depth. 

The mean effective temperature of the stars in the TEPCat sample is $\left< T_{\rm eff}\right> \simeq 5850$ K. We generated a mock catalogue with a temperature-dependent tidal quality factor of the form
\begin{equation}
\log_{10} Q_s'(T_{\rm eff})=\log_{10} Q_s'(5850\ {\rm K})+ \epsilon(T_{\rm eff}-5850)/1000,
\end{equation}
with $\log_{10} Q_s'(5850\ {\rm K})=8.3$ and $\epsilon = -0.7$, 0.0 and +0.7. Two-dimensional plots of the relative likelihood are given in Fig.~\ref{fig:mocklikely2D}. The maximum-likelihood estimates of $\log_{10}Q_s'$ and $\epsilon$  recover the input values with reasonable fidelity. A mock catalogue constructed with $\{\log Q,\epsilon\} =\{8.3,-0.7\}$ yielded $\{8.16\pm 0.17,-0.61\pm 0.26\}$, while $\{\log Q,\epsilon\} =\{8.3,+0.7\}$ returned $\{8.101\pm 0.21,0.61\pm 0.22\}$. 

\begin{figure*}
\includegraphics[width=\columnwidth]{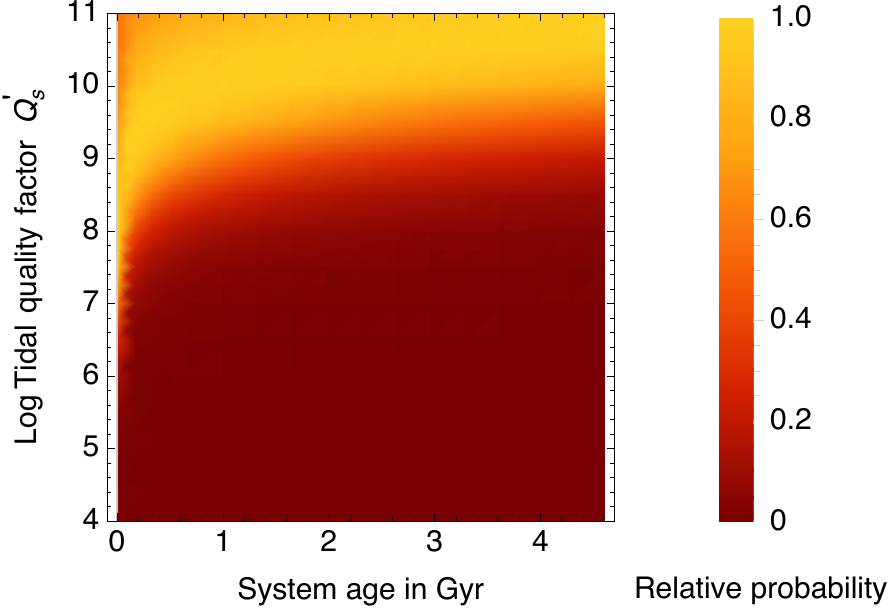}
\includegraphics[width=\columnwidth]{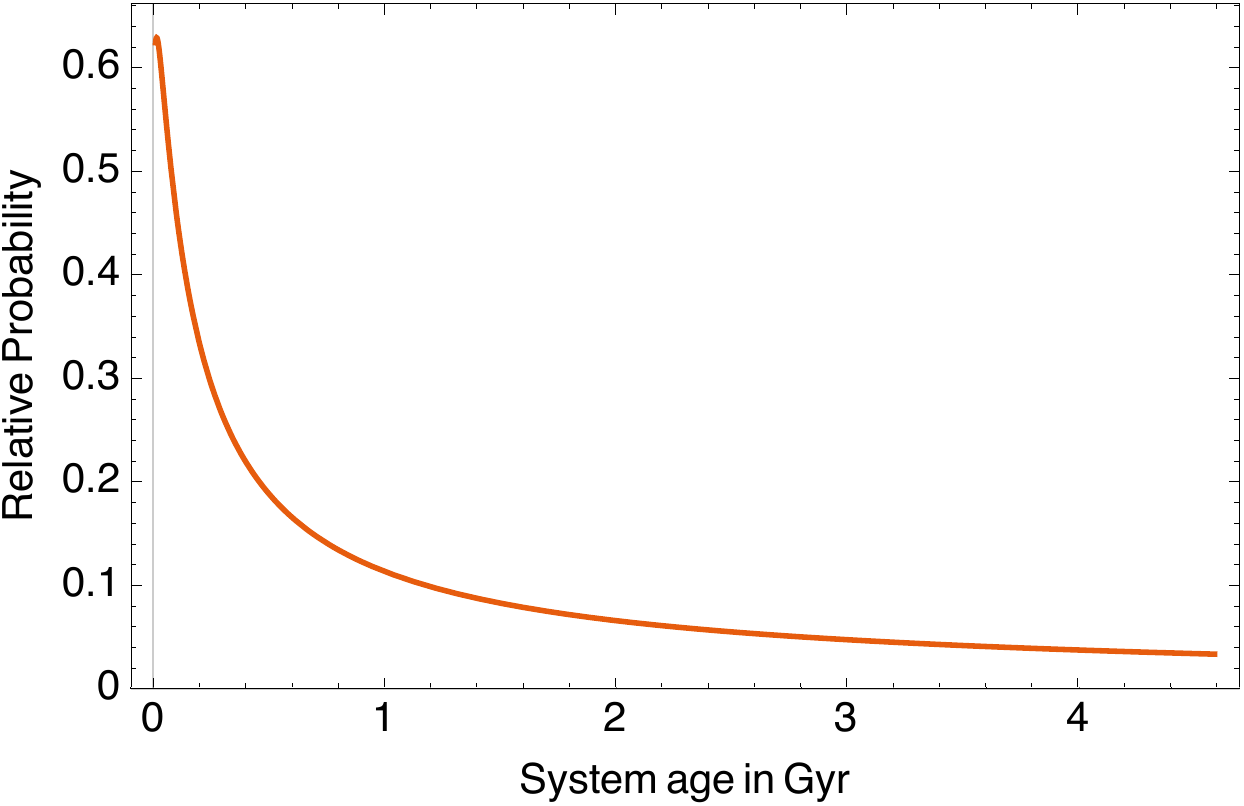}
\includegraphics[width=\columnwidth]{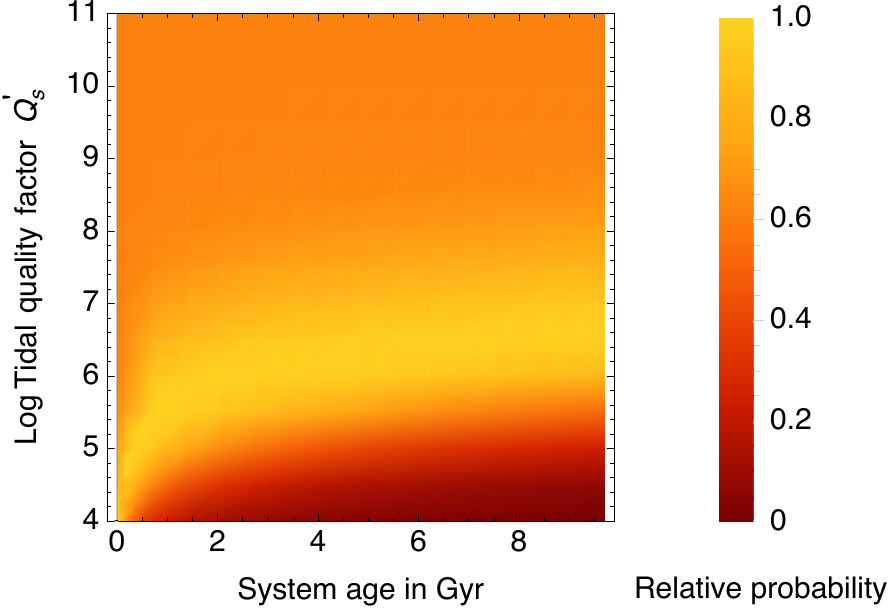}
\includegraphics[width=\columnwidth]{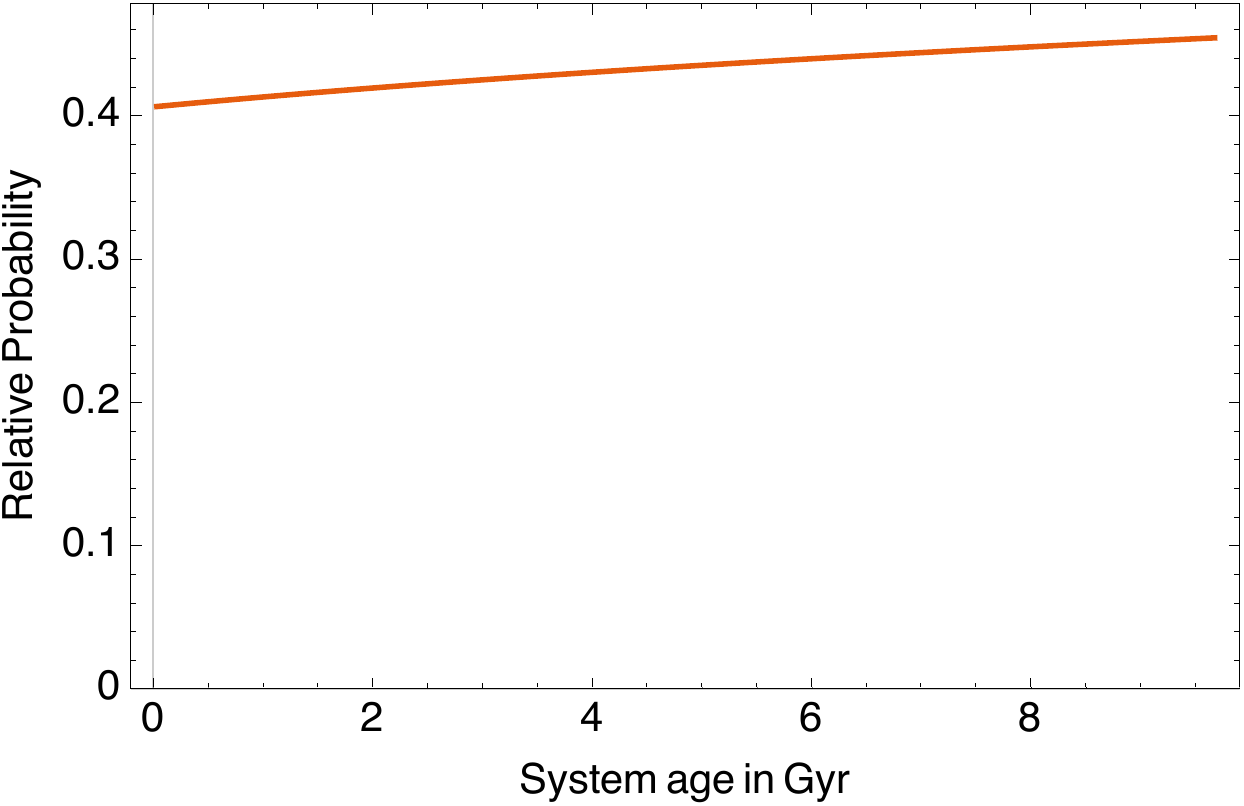}
\caption{Variation in probability density with age and tidal quality factor for WASP-18b (upper) and WASP-21b (lower). In the left-hand panels, the ridge line of maximum probability curves downward at young ages. In the right-hand panels, a cut through the distribution at $\log Q_s'=8.26$ favours a young age for WASP-18b, but leaves the age of WASP-21b indeterminate.}
\label{fig:ageQ}
\end{figure*}

\section{The TEPCat sample: $\log Q_s'=8.3$}
\label{sec:tepcat}

Having verified that the tidal quality factor and its temperature dependence can be inferred accurately at population level, we now apply the method outlined in Section~\ref{sec:probdens} to a comparable sample of real systems. The system parameters were taken from the TEPCat catalogue of well-studied transiting planets \citep{2011MNRAS.417.2166S}. We included all 242 transiting planets detected in the WASP, HAT/HAT-S, TrES, XO, QES, KELT, MASCARA and WTS surveys, with in the mass range $0.3 < M_p/M_{\rm Jup}<13.0$. 

For most of these planets, the orbital period $P_{\rm orb}<<P_{\rm rot}$, the stellar rotation period. For many of the TEPCat systems, the stellar rotation period can be estimated from the stellar radius and $v\sin i$. Using $v\sin i$ and projected-obliquity values drawn from the the Exoplanet Orbit Database \citep{2014PASP..126..827H}, we identified a subset of 25 systems for which $0.5<P_{\rm orb}/P_{\rm rot}<2.0$. In this period-ratio range, \cite{2007ApJ...661.1180O} predict that the dynamical tide will excite inertial waves in the stellar interior, enhancing the tidal coupling strength. Among these 25 systems, 6 have projected obliquities in the range $80^\circ < \lambda < 180^\circ$. The misaligned and/or retrograde motion of these six planets raises their tidal forcing frequencies well above the values suggested by a simple period ratio. They and the remaining 223 systems are likely to lie in the equilibrium-tide regime of \cite{2007ApJ...661.1180O}. The 19 prograde planets with $0.5 < P_{\rm orb}/P_{\rm rot}< 2.0$ are HAT-P-2b, 9b, 24b, 33b, 34b, 39b, 41b; KELT-7b; TrES-1b; WASP-38b, 59b, 61b, 62b, 66b, 84b, 101b and 106b; X0-2Sb and 3b. The six systems with low forcing frequencies but highly-inclined or retrograde orbits are HAT-P-14b and 32b; WASP-7b, 17b, 33b and 79b.

The resulting scatter plots, and likelihoods $\mathcal{L}(\boldsymbol{x}|Q_s',\epsilon)$ are plotted in Fig.~\ref{fig:realdata} for all three subgroups. The inferred values of the tidal quality factor are listed in Table~\ref{tab:TEPCatvalues}. The results confirm strongly that the 19 prograde systems with $0.5 < P_{\rm rot}/P_{\rm orb} < 2.0 $ have a collective value of $\log Q_s'=7.3\pm0.4$. This is an order of magnitude lower than the $\log Q_s'=8.3\pm0.14$ of the remaining 223 systems. This finding is consistent with the enhancement in tidal coupling strength expected in the regime where the dynamical tide becomes important \citep{2007ApJ...661.1180O}. 

The left-hand panels of Figure~\ref{fig:realdata} show the wide scatter of points seen in the mock catalogues generated with a uniform distribution in $\log P_{\rm init}$, though there is also some evidence of a pile-up at about 3 times the Roche limit. This pattern is suggestive of a mixture of eccentric and disc migration, as discussed in Section~\ref{sec:retrievaltests} above.

\begin{table}
\caption{Results of parameter retrieval tests derived from the TEPCat sample. 
The first column identifies the subset;
the second column gives the number of systems in the subset; and the third
column gives the inferred values of $Q_s'$.
}
\label{tab:TEPCatvalues}
\begin{tabular}{ccc}
Subsample & $N_{\rm sys}$ & $\log Q_s'$ \\
\hline\\
Prograde: $0.5 < P_{\rm rot}/P_{\rm orb} < 2.0 $ & 19 & $7.31 \pm 0.39$ \\
Inclined/Retrograde: $0.5 < P_{\rm rot}/P_{\rm orb} < 2.0$ & 6 & $8.33\pm 1.11$\\
Full sample excluding prograde above & 223 & $8.26 \pm 0.14$\\
Full sample & 242 & $8.19\pm 0.14$\\
\hline\\
\end{tabular}
\end{table}

The probability map for the temperature-dependent $Q_s'$ model shown in Fig.~\ref{fig:likely2D} reveals no significant dependence of $Q_s'$ on effective temperature $T_{\rm eff}$, with $\{\log Q,\epsilon\} =\{8.24\pm 0.19,+0.17\pm 0.25\}$. A positive temperature dependence might be expected from theoretical predictions \citep{2009MNRAS.395.2268B,2010MNRAS.404.1849B,2011ApJ...731...67P,2015A&A...580L...3M} that tidal dissipation rates are enhanced in stars with deeper convective zones. If such a dependence is present, it is too weak, with a 2-$\sigma$ upper limit $\epsilon < 0.67$, to detect in the present sample.

\subsection{Individual ages and spiral-in times}
\label{sec:ages}

One of the most powerful features of hierarchical Bayesian inference is that, once the population-level hyperparameters have been determined, they can be used to infer the values of quantities that were previously treated as unconstrained nuisance parameters. In this application, the stellar age is such a parameter. 

To infer the population mean value of $Q_s'$ from a sample of stars of largely unknown (or at best weakly-constrained) ages, we marginalised over the stellar age using a uniform prior. With the value of $Q_s'$ now being reasonably tightly constrained, we can compute the probability density $p(x_j|k(\boldsymbol{\theta}_j, Q_s', \epsilon), t_{\rm age,j})$ as a function of stellar age, normalising eq.~\ref{eq:rhoxt} by integrating over $x$ from $x_{\rm min}$ to a sufficiently large value of $x$ to capture most of the probability. We should, strictly, marginalise $p(x_j|k(\boldsymbol{\theta}_j, Q_s', \epsilon), t_{\rm age,j})$ over the new distribution inferred for $Q_s$'. Inspection of Fig.~\ref{fig:ageQ} shows, however, that the length scale of variation with respect to $Q_s'$ is large compared with the uncertainty in $Q$. To keep the problem simple and analytic, we approximate the new prior for $Q_s'$ as a delta function at the most likely value, i.e $\log Q_s'=8.26$, with no temperature dependence ($\epsilon=0.0$). The normalising integral of eq.~\ref{eq:rhoxt} is once again lengthy but at least analytic and real.

The normalised probability densities are shown as a function of age $t$ and $Q_s'$ in Fig.~\ref{fig:ageQ} for WASP-18b and WASP-21b. Although the probability is not a strong function of age, a cut through the figure at the globally determined value of $Qs'$ shows that for WASP-18b, which has the shortest spiral-in time in the sample, the probability density at is greatest at small ages. 

In principle we could use this new information to construct an improved prior on stellar age, by normalising the curves shown in the lower panels of Fig.~\ref{fig:ageQ}, and using them to replace the uninformative prior $\pi{t}$. To discover whether recalculating the likelihoods with improved age information would reduce the uncertainty in $Q_s'$ significantly, we recalculated the likelihood as a function of $Q_s'$ for the mock planet catalogues, the ages of whose individual stars are known exactly. This entailed replacing the uninformative prior on each star's age with a delta function at the known age, and recalculating the likelihood using eq.~\ref{eq:rhoxt} in place of eq.~\ref{eq:rhox}. The resulting likelihood function was so nearly identical in position and width to that obtained using the original procedure, that we elected not to iterate.

The inferred age distributions for individual stars are nonetheless useful in understanding some previously-unexplained features of the hot-Jupiter population. As eq.~\ref{eq:rhoxt} shows,  planets close to their stars retain probability densities close to the primordial value when they are young enough that $13 k t/2 << e^{13 x/2}$. An older planet at the same present-day position has migrated further and speeded up since birth, so its probability density at the same location is lower. Although the WASP-18 star has a main-sequence lifetime of nearly 4.6 Gyr, half the cumulative probability lies at ages less than 1.1 Gyr. Similarly, the ``median age" of WASP-19b is 2.4 Gyr, even though its host star's nuclear lifetime is greater than the age of the galactic disk.

This bias in favour of younger ages may explain why the host stars of many of the most rapidly-decaying planets such as WASP-18 and WASP-19 have rotational ages that are significantly shorter than their main-sequence lifetimes \citep{2011MNRAS.415..605B}. It cannot, however, explain why some hot-Jupiter hosts appear to have gyrochronological ages significantly smaller than their isochronal ages \citep{2014MNRAS.442.1844B,2015A&A...577A..90M}.

The less extreme hot Jupiters, such as the less massive and more distant WASP-21b, show very little variation in probability density with age, as seen in the lower-right panel of Fig.~\ref{fig:ageQ}. Even in the most dramatic cases such as WASP-18b, we caution that the variation in probability density with age is insufficient to serve as a precise age indicator. 

With the tidal quality factor fixed at a temperature-independent value $\log Q_s'=8.26$, only a few of the shortest spiral-in times in the sample are under 1 Gyr. WASP-18b will be disrupted at its Roche limit in 0.09 Gyr. KELT-16b (0.20 Gyr) and HATS-18b (0.39 Gyr) will suffer similar fates. The lower-density WASP-103b (0.12 Gyr), WASP-19b (0.14 Gyr),  WASP-12b (0.13 Gyr)  and  KELT-09b (0.54 Gyr) may undergo hydrodynamic envelope loss before reaching the Roche limit.

\subsection{Tidal age bias, planetary surface gravities and chromospheric activity.}

The inverse relationship between probability density and inward drift rate expressed in Eq.~\ref{eq:rhoxt}, and the resulting bias toward young ages for the most rapidly-decaying planets,  provides a statistical  explanation for the apparently anomalous correlation between the chromospheric emission fluxes of hot-Jupiter hosts and the surface gravities of their close-orbiting planets. This correlation, initially reported by \cite{2010ApJ...717L.138H} and subsequently confirmed by \cite{2014A&A...572A..51F}, appears to be statistically robust \citep{2016OLEB...46..385F}, but has so far defied physical explanation. 

Surface gravity correlates strongly with planet mass, and the tidal migration rate scales with the square of planet mass in eq.~\ref{eq:dlnadt}.  Any planet whose life expectancy is significantly shorter than its main-sequence lifetime is more likely to be seen when it is young, if our assumption that the planetary birth rate  is more or uniform in log separation is anywhere near correct. The probability density at birth is attenuated by the accelerating inward drift rate as the star ages. The relative probability density of seeing the planet as it passes its present location depends on how rapidly it is migrating inwards, and is given directly by Eq.~\ref{eq:rhoxt}.

\begin{figure}
\includegraphics[width=\columnwidth]{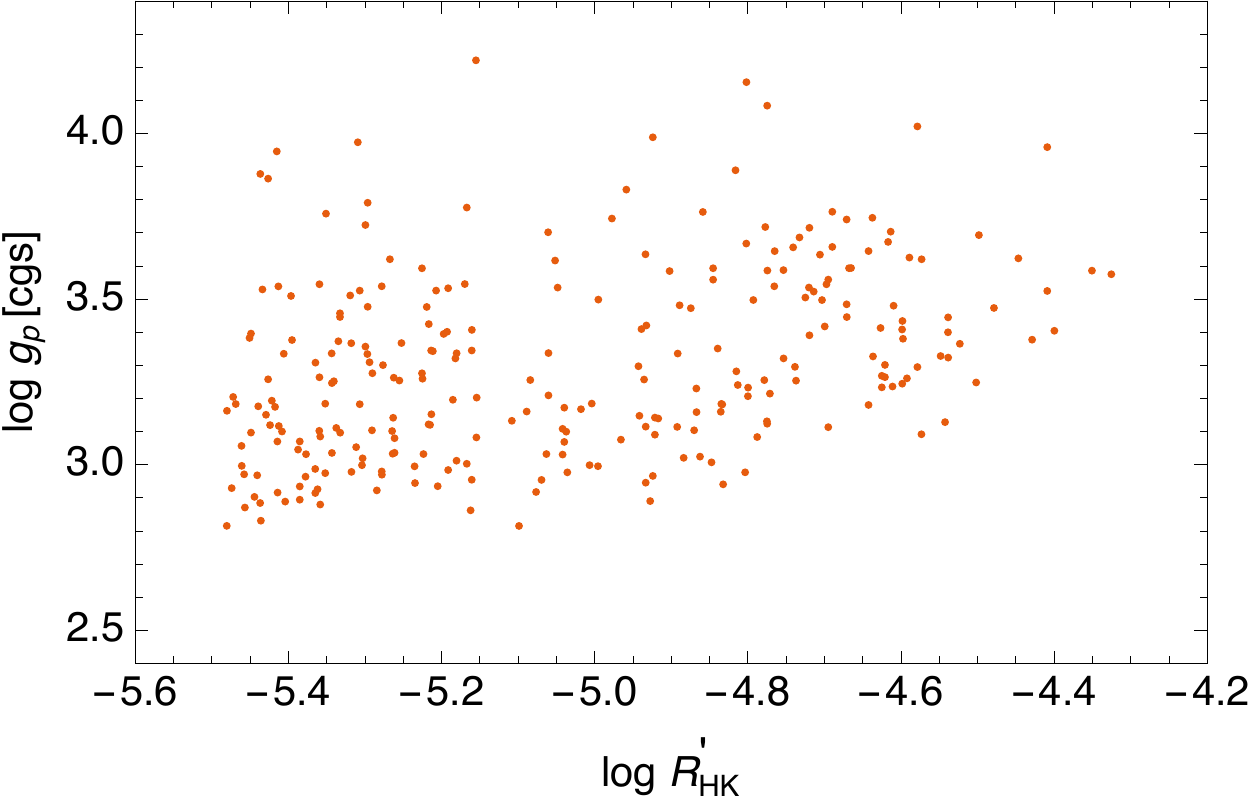}
\caption{Planetary surface gravity versus chromospheric $R'_{HK}$ for a mock catalogue of transiting hot Jupiters generated with 
$\log Q_s'$=8.26 and $\epsilon=0.00$ (i.e. no temperature dependence). 
The correlation appears to arise from the statistical bias towards younger systems produced by the faster inward migration of more massive planets.}
\label{fig:rhkgpplot}
\end{figure}

To test this idea we used the calibration of \cite{2007ApJ...669.1167B} to estimate stellar rotation periods $P$ from the $B-V$ colour indices and main-sequence ages of stars in the TRILEGAL mock catalogue. We used the calibrations of \cite{1984ApJ...279..763N} to estimate convective turnover times $\tau_c$ from $B-V$ and to estimate the chromospheric $\log R'_{HK}$ from $\log(P/\tau_c)$. The surface gravities of the mock catalogue of planets, generated with 
$\log Q_s'$=8.26 and $\epsilon=0.00$, 
are plotted against $\log R'_{HK}$ in Fig.~\ref{fig:rhkgpplot}, using the same axis ranges as Fig 1 of \cite{2010ApJ...717L.138H}, and similar to those of Fig. 1 of \cite{2014A&A...572A..51F}. We see a similarly broad correlation, with a lower boundary sharing almost the same location and slope as Hartman's and Figueira's plots. The correlation is clearly real, and is equally clearly  a selection effect resulting from tidal age bias. The mock catalogue incorporates no star-planet interaction physics.

\subsection{Transit timing variations}

Short though they may be relative to stellar lifetimes, the spiral-in times inferred in Section~\ref{sec:ages} are long in human terms. A more practical consideration is the length of time we must wait before the decreasing orbital period produces a  departure from a linear transit ephemeris great enough to be detectable in a long-term transit-timing campaign. 

The orbital distance of a planet initially located at $a_0$ decreases with time $t$ as
\begin{equation}
a=a_0\left(1-\frac{13}{2}\frac{k t}{(a_0/R_s)^{13/2}}\right)^{2/13}
\end{equation}
Kepler's third law gives the orbital angular frequency 
\begin{equation}
n=\sqrt{\frac{G M_s}{a^3}}=\sqrt{\frac{G M_s}{a_0^3}}\left(1-\frac{13}{2}\frac{k t}{(a_0/R_s)^{13/2}}\right)^{-3/13}.
\end{equation}

\begin{figure}
\includegraphics[width=\columnwidth]{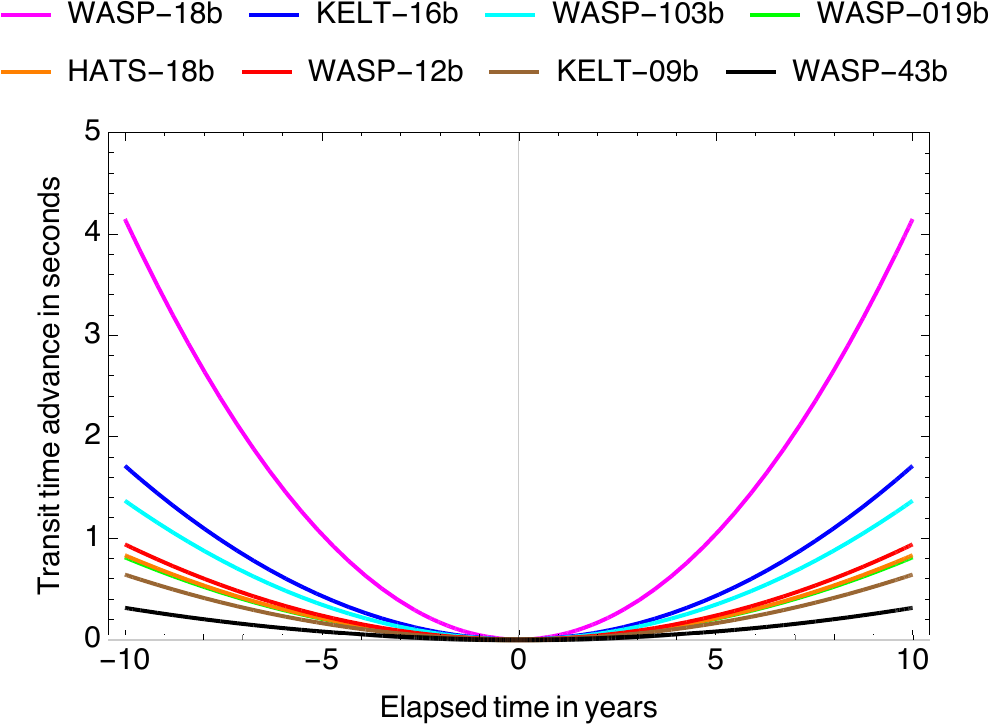}
\caption{
Transit timing advance over a 20-year baseline of elapsed time for the eight most rapidly-decaying hot Jupiters in the TEPCat sample, using $\log Q_s'=8.26$ and $\epsilon=0$. 
}
\label{fig:delayplot}
\end{figure}

The total angle swept out in time $t$ is found by using $n = d\theta/dt$ and integrating
\begin{equation}
\theta(t)=\sqrt{\frac{G M_s}{a_0^3}}
\left(\frac{(a_0/R_s)^{13/2}}{5 k} \left(1-\left(1-\frac{13}{2}\frac{k t}{
   (a_0/R_s)^{13/2}}\right)^{10/13}\right)\right).
\label{eq:theta}
\end{equation}

The advance in transit time from the constant-period ephemeris is 
\begin{equation}
\frac{\delta\theta(t)}{n}
=\frac{(a_0/R_s)^{13/2}}{5 k} \left(1-\left(1-\frac{13}{2}\frac{k t}{
   (a_0/R_s)^{13/2}}\right)^{10/13}\right)-t.
\label{eq:deltat}
\end{equation}
This is a sufficiently small difference between large quantities that for computational purposes eq.~\ref{eq:deltat} is best implemented as a series expansion:
\begin{equation}
\delta t=\frac{3 k t^2}{4 r_0^{13/2}}+\frac{2 k^2 t^3}{r_0^{13}}+\frac{29 k^3 t^4}{4r_0^{39/2}}+\frac{609 k^4 t^5}{20 r_0^{26}}+\mathcal{O}\left(t^6\right).
\end{equation}
Only the first term is needed over timescales of interest to human observers. The same expression is obtained via the Taylor-series approach to the same problem by \cite{2014MNRAS.440.1470B}. 

The resulting quadratic models for the expected variations in observed-minus-computed (O-C) transit times are illustrated in Fig.~\ref{fig:delayplot} for the eight most rapidly-decaying planets in the sample, using the most likely value $Q_s'=10^{8.26}$ inferred from the period-separation diagram. Given that the typical precision of the best transit-timing observations of hot Jupiters is of order a few seconds when averaged over all transits in a single observing season, the prospects for detecting the orbital acceleration of WASP-18 appear marginally achievable on timescales of 1 to 2 decades. The measurement will be sufficiently challenging that high-cadence space observations from space-photometry missions such as NASA's {\em TESS} or the Swiss-led ESA {\em CHEOPS} mission are likely to be needed.

\section{Discussion and Conclusions}

The sharp upper-left boundary in the mass-separation diagram for hot Jupiters lies along a contour of constant tidal spiral-in time, as calculated using equilibrium tide theory. In this paper we have applied a form of Bayesian hierarchical inference to a set of mock datasets with known properties and sizes similar to the presently-known catalogue of hot Jupiters. Our results show that, if a universal tidal quality factor $Q_s'$ exists amongst planet-host stars whose planets impose tidal forcing frequencies high enough for equilibrium tide theory to be applicable, the value of $Q_s'$ can be determined within a factor 2 over a range of values spanning at least 5 orders of magnitude. We find evidence of even stronger tidal depletion, among a sub-sample of 19 prograde systems whose low tidal-forcing frequencies plausibly place them in the dynamical-tide regime. 

Among the 223 equilibrium-tide systems in our sample, we find no significant dependence of the tidal quality factor on stellar effective temperature. We find that 
$\log Q_s'=8.26\pm 0.14$. Among the 19 systems with prograde orbits for which the ratio of the stellar rotation period to the planet's orbital period lies in the range $0.5 < P_{\rm rot}/P_{\rm orb}<2.0$, we find evidence of tidal coupling an order of magnitude stronger, with $\log Q_s'=7.31\pm 0.39$. The low tidal forcing frequencies of these systems plausibly put them in the regime where the dynamical tide may excite inertial waves in the stellar interiors, leading to stronger dissipation.

These results may not be as accurate as they are precise. What could possibly go wrong? The inverse relationship between probability density and the local rate of inward drift is sound and consistent with the Hertzsprung-gap analogy.  

Our assumption that newborn hot Jupiters appear at locations with a flat distribution in log separation is consistent with the flat distribution in $\log P$ out to periods of a year or so that has emerged from radial-velocity surveys \citep{2007ARA&A..45..397U}.
Nonetheless, we note that the mass-separation diagram shows some evidence of a pile-up of planets just outside twice their Roche limits, as might be expected if a fraction of the sample had undergone eccentric migration followed by tidal circularisation. 
Having also validated our retrieval algorithm using mock catalogues generated with a gaussian (rather than flat) initial distribution in log period, we are confident that the method is robust for such non-flat distributions over the range $6.5 < \log Qs'<8.5$. 

Our analytic expression for the probability density function in log separation includes the inverse relationship between scaled separation $a/R_s$ and transit probability, but does not account for the detection bias suffered by real transit surveys and mimicked in our mock catalogues.  

The method succeeds nonetheless in retrieving the input values of $Q_s'$ and $\epsilon$ accurately from the mock catalogues.  We attribute this robustness to the fact that the planets which constrain the hyperparameters most severely are those in the closest orbits, for which sample incompleteness is far less of an issue that it is for more distant planets.

Having established such tight constraints on the population value of $Q_s'$, we have been able to recover probability density functions for the ages of individual systems. The key insight gained is that we are far more likely to see young planets than old ones, in locations close enough to their stars to have spiral-in times significantly shorter than the star's nuclear lifetime. This is another straightforward consequence of the inverse relationship between probability density and drift rate. This hitherto unrecognised selection effect provides a subtle but convincing explanation for the statistically robust but otherwise mysterious correlation between planetary surface gravity and stellar chromospheric activity levels. This success also provides a test of our assumption of a flat distribution in initial periods is consistent with reality. The selection effect would take a different form if the initial period distribution had a strongly non-uniform dependence on log period.

Finally, knowledge of $Q_s'$ allows us to forecast how long it will take to detect a departure from a linear ephemeris in the transit times of WASP-18, the fastest decaying planet known. 
With a 4-second departure from a linear ephemeris over a 20-year baseline, the predictions are at best challenging, even for the  next decade of space-borne photometry. With luck, the forthcoming {\em TESS} mission will identify new close-in hot Jupiters with decay rates even greater than that of WASP-18. These should become prime targets for transit-timing studies with the {\em CHEOPS} mission. The population-level probability distributions shown in Figure 9, properly normalised, will be useful as priors for Bayesian determinations of $Q_s'$ for these individual systems.

\section*{Acknowledgements}

We are grateful to the anonymous referee for insightful comments which led to several improvements including the discovery of the enhanced tidal coupling in the dynamical-tide regime. ACC and MMJ acknowledge support from STFC Consolidated Grant ST/M001296/1. This research has made use of the Exoplanet Orbit Database and the Exoplanet Data Explorer at \verb|http://exoplanets.org| and the TEPCat Database at \verb|http://www.astro.keele.ac.uk/jkt/tepcat/tepcat.html|. 








\appendix

\section{Access to underpinning data}

This work is based on a subset of stellar and planetary data compiled from the TEPCat and Exoplanet Orbit databases, as described in Section~\ref{sec:tepcat}. Machine-readable, merged tables of the important stellar and planetary measurements and derived parameters that contributed to the analysis are available, together with other supplementary tables, at \verb|http://dx.doi.org/10.17630/f449bc99-ccaa-45ff-9142-3cca49f7c786|.


\bsp	
\label{lastpage}
\end{document}